\documentclass[pra,twocolumn,showkeys]{revtex4}
\usepackage{amsmath,bm,verbatim}
\usepackage{fleqn}

\renewcommand{\a}{\mathbf{a}}

\newcommand{\B}{\mathbf{B}}
\renewcommand{\b}{\mathbf{b}}
\renewcommand{\d}{\mathbf{d}}
\newcommand{\E}{\mathbf{E}}
\newcommand{\e}{\mathbf{e}}
\renewcommand{\i}{\mathbf{i}}
\newcommand{\p}{\mathbf{p}}
\renewcommand{\r}{\mathbf{r}}
\newcommand{\s}{\mathbf{s}}
\renewcommand{\u}{\mathbf{u}}
\renewcommand{\v}{\mathbf{v}}
\newcommand{\z}{\mathbf{z}}
\newcommand{\x}{\mathbf{x}}
\newcommand{\obe}{\overline e}
\newcommand{\obF}{\overline F}
\newcommand{\obh}{\overline h}
\newcommand{\obm}{\overline m}
\newcommand{\obS}{\overline S}
\newcommand{\obp}{\overline p}
\newcommand{\obu}{\overline u}
\newcommand{\obv}{\overline v}

\newcommand{\calM}{{\mathcal M}}

\newcommand{\talpha}{\widetilde\alpha}
\newcommand{\tpsi}{\widetilde\psi}

\newcommand{\ta}{\widetilde a}
\newcommand{\tF}{\widetilde F}
\newcommand{\tL}{\widetilde L}
\newcommand{\tN}{\widetilde N}
\newcommand{\tM}{\widetilde M}
\newcommand{\tR}{\widetilde R}
\newcommand{\tU}{\widetilde U}

\renewcommand{\AA}{\hbox{$\mathring{\textrm{A}}$}}

\newcommand{\btheta}{\bm{\theta}}
\newcommand{\bsig}{\bm{\sigma}}
\newcommand{\bmu}{\bm{\mu}}
\newcommand{\bnabla}{\bm{\nabla}}

\newcommand{\half}{{\textstyle \frac{1}{2}}}
\newcommand{\smallhalf}{{\scriptstyle\frac{1}{2}}}
\newcommand{\bdot}{\bm{\cdot}}
\newcommand{\btimes}{\bm\times}

\begin{document}

\title{Zitterbewegung in Quantum Mechanics -- a research program}

\author{David Hestenes}
\affiliation{Department of Physics, Arizona State University, Tempe, Arizona 85287-1504}
\email{hestenes@asu.edu}
\homepage{http://modelingnts.la.asu.edu/}

\begin{abstract}
Spacetime Algebra (STA) provides unified, matrix-free spinor methods for rotational
dynamics in classical theory as well as quantum mechanics.
That makes it an ideal tool for studying particle models of \textit{zitterbewegung}
and using them to study zitterbewegung in the Dirac theory. This paper develops a
self-contained dynamical model of the electron as a lightlike particle with helical
zitterbewegung and electromagnetic interactions.
It attributes to the electron an electric dipole moment oscillating with ultrahigh
frequency, and the possibility of observing this directly as a resonance in electron
channeling is analyzed in detail.
A modification of the Dirac equation is suggested to incorporate the oscillating
dipole moment. That enables extension of the Dirac equation to incorporate
electroweak interactions in a new way.
\end{abstract}


\maketitle

\section{Introduction}

This paper continues a research program investigating implications of
the  \textit{Real Dirac Equation} for the interpretation and extension of quantum
mechanics. Details of the program have been reviewed elsewhere
\cite{Hest03b,Hest03c,Hest96}, so it suffices here to state the main ideas and
conclusions to set the stage for the present study.

The program began with a reformulation of the Dirac equation in terms of \textit{Spacetime
Algebra} (Section \ref{secII}), which revealed geometric structure that is suppressed in
the standard matrix version. In particular, it revealed that the generator of
phase and electromagnetic gauge transformations is a spacelike bivector
specified by electron spin. In other words, spin and phase are inseparably related
--- spin is not simply an add-on, but an essential feature of quantum mechanics.
However, physical implications of this fact depend critically on relations of the
Dirac wave function to physical observables, which are not specified by the Dirac
equation itself. That started the present research program to investigate various
possibilities.

A standard observable in Dirac theory is the Dirac current, which doubles as
a probability current and a charge current. However, this does not account for the
magnetic moment of the electron, which many investigators conjecture is due to a
circulation of charge. But what is the nature of this circulation? After a lengthy
analysis of the Dirac equation Bohm and Hiley conclude \cite{Bohm93}: ``the electron
must still be regarded as a simple point particle whose only intrinsic property is
its position.''  Under this assumption, spin and phase must be kinematical features
of electron motion. The charge circulation that generates the magnetic moment can
then be identified with the \textit{zitterbewegung} of Schroedinger \cite{Schr30}.

This raises the central question of the present research: Is the zitterbewegung, so
construed, a real physical phenomenon, or is it merely a colorful metaphor?
Although this question was motivated by structural features of the Dirac equation, it
cannot be answered without attributing substructure to electron motion that is not
specified by standard Dirac theory.

The main purpose of this paper is to formulate and study a well-defined particle model
of the electron with \textit{spin and zitterbewegung dynamics} motivated by the
Dirac equation. Since the term \textit{zitterbewegung} is quite a mouthful, I often
abbreviate it to \textit{zitter}, especially when it is used as an adjective.

We study the structure of the \textit{zitter model} in considerable detail
with the aim of identifying new experimental implications. The main conclusion is
that the electron is the seat of a rapidly rotating electric dipole moment
fluctuating with the zitter frequency of Schroedinger. As this frequency is so rapid,
it is observable only under resonance conditions. It is argued that many
familiar quantum mechanical effects may be attributable to zitter resonance. Moreover, the
new possibility of observing zitter directly as a resonance in electron channeling is
analyzed in detail, because the prospects of crucial experimental tests are very
promising.

The relation of the zitter particle model to the Dirac equation is also studied. The main
conclusion is that, though zitter oscillations are inherent in the Dirac equation, they
will not be manifested as an oscillating electric dipole  without altering the definition
of charge current. A simple modification of the Dirac equation to incorporate the altered
definition is proposed.  Remarkably, that opens the door for incorporating electroweak
interactions into the Dirac equation in a novel way.

In conclusion, the relation of the zitter particle model to the Dirac equation can be
considered from two different perspectives.
On the one hand, it can be regarded as a ``quasiclassical" approximation that embodies
structural features of the Dirac equation in a convenient form for analysis.
On the other hand, it can be regarded as formulating fundamental properties of the
electron that are manifested in the Dirac equation in some kind of average form.
The choice of perspective is left to the reader.

\section{Spacetime Algebra}\label{secII}

Spacetime algebra is thoroughly expounded elsewhere \cite{Hest03b}, so a brief
description is sufficient here, mainly to define terms.

We represent Minkowski \textit{spacetime} as a real 4-dimensional vector
space $\calM^4$. The two properties of scalar multiplication and vector
addition in $\calM^4$ provide only a partial specification of spacetime
geometry.  To complete the specification we introduce an associative
\textit{geometric product} among vectors $a, b, c,\ldots$ with the property
that the square of any vector is a (real) scalar. Thus for any vector $a$
we can write
\begin{equation}
a^2 = a a  = \epsilon|a|^2\,,\label{2.1}
\end{equation}
where $\epsilon$ is the \textit{signature} of $a$ and $|a|$ is a (real)
positive scalar. As usual, we say that $a$ is \textit{timelike},
\textit{lightlike} or \textit{spacelike} if its signature is positive
($\epsilon = 1$), null ($\epsilon = 0$), or negative ($\epsilon =-1$).
We can specify the signature of $\calM^4$ as a whole, by adopting the
axioms: (a) $\calM^4$ contains at least one timelike vector; and (b)
every 2-plane in $\calM^4$ contains at least one spacelike vector.

To facilitate applications of STA to physics a few definitions
and theorems are needed. From the \textit{geometric product} $uv$
of two vectors it is convenient to define two other
products. The \textit{inner product} $u\bdot v$ is defined by
\begin{equation}
u\bdot v = \half(uv + vu)=v\bdot u\,,\label{2.2}
\end{equation}
while the \textit{outer product } $u\wedge v$ is defined by
\begin{equation}
u\wedge v = \half(uv - vu) = - v\wedge u\,.\label{2.3}
\end{equation}
The three products are therefore related by
\begin{equation}
uv =u\bdot v + u\wedge v\,.\label{2.4}
\end{equation}
This can be regarded as a decomposition of the
product\ $uv$ into symmetric and skewsymmetric parts, or
alternatively, into scalar and bivector parts.

The inner and outer products\ can be generalized.
One way is to define the outer product\ along
with the notion of $k$-\textit{vector} iteratively as
follows: scalars are defined to be 0-vectors,
vectors are 1-vectors, and bivectors, such as $u\wedge v$,
are 2-vectors. For a given $k$-vector $K$, the
integer $k$ is called the \textit{grade} of $K$.
For $k\ge1$, the outer product\ of a
vector $v$ with a $k$-vector $K$ is a $(k+1)$-vector
defined in terms of the geometric product\ by
\begin{equation}
v\wedge K =\half(vK+(-1)^kKv)=(-1)^kK\wedge v\,.\label{2.5}
\end{equation}
The corresponding inner product\ is defined by
\begin{equation}
v\bdot K =\half(vK+(-1)^{k+1}Kv)=(-1)^{k+1}K\bdot v\,,\label{2.6}
\end{equation}
and it can be proved that the result is a $(k-1)$-vector.
Adding (\ref{2.5}) and (\ref{2.6}) we obtain
\begin{equation}
vK=v\bdot K + v\wedge K\,,\label{2.7}
\end{equation}
which obviously generalizes (\ref{2.4}). The important thing
about (\ref{2.7}), is that it decomposes $vK$
into $(k-1)$-vector and $(k+1)$-vector parts.

\def\g{\hbox{g}}
By continuing as above, STA as been developed into a complete
coor\-di\-nate-free calculus for spacetime physics \cite{Hest03b}.
However, to hasten comparison with standard Dirac algebra, we
interrupt that approach to introduce coordinates and a
basis for the algebra. Let $\{\gamma_\mu;\ 0,1,2,3\}$
be a \textit{right-handed orthonormal frame} of vectors
with $\gamma_0$  in the forward light cone. In
accordance with (\ref{2.2}), the
components $\g_{\mu\nu}$   of the metric tensor
for this frame are given by
\begin{equation}
\g_{\mu\nu}=\gamma_\mu\bdot\gamma_\nu
=\half(\gamma_\mu\gamma_\nu+\gamma_\nu\gamma_\mu)\,.\label{2.8}
\end{equation}
This will be recognized as isomorphic to a famous formula
of Dirac's. The difference here is that the $\gamma_\mu$
are vectors rather than matrices.  A \textit{coframe} $\{\gamma^\mu\}$ is
defined by the usual convention for raising and lowering indices:
$\gamma_\mu=\g_{\mu\nu} \gamma^\nu$.

The unit pseudoscalar $i$  for spacetime is related
to the frame $\{\gamma_\nu\}$ by the equation
\begin{equation}
i=\gamma_0\gamma_1\gamma_2\gamma_3
=\gamma_0\wedge\gamma_1\wedge\gamma_2\wedge\gamma_3\,.\label{2.9}
\end{equation}
It is readily verified from (\ref{2.9}) that $i^2=-1$,
and the geometric product\ of $i$ with any vector is
anticommutative.

By multiplication the $\gamma_\mu$ generate a
complete basis of $k$-vectors for STA, consisting
of the $2^4 = 16$ linearly independent elements
\begin{equation}
1,\quad\gamma_\mu,\quad\gamma_\mu
    \wedge\gamma_\nu,\quad\gamma_\mu i,\quad i\,.\label{2.10}
\end{equation}
Multivectors with even grade compose a subalgebra of the STA
generated by the bivectors $\{\bsig_k \equiv \gamma_k\gamma_0; k= 1,2,3\}$, so that
\begin{equation}
\bsig_1\bsig_2\bsig_3= \gamma_0\gamma_1\gamma_2\gamma_3=i.\label{2.11}
\end{equation}
Any multivector can be expressed as a linear combination of these elements.

For example, a bivector $F$ has the expansion
\begin{equation}
F = \half F^{\mu\nu}\gamma_\mu\wedge\gamma_\nu
,\label{2.12}
\end{equation}
with its ``scalar components" $F^{\mu\nu}$ given by
\begin{equation}
F^{\mu\nu}=\gamma^\mu\bdot F\bdot\gamma^\nu
         =\gamma^\nu\bdot( \gamma^\mu\bdot F)
    = (\gamma^\nu\wedge\gamma^\mu)\bdot F\,.\label{2.13}
\end{equation}
Note that the two inner products\ in the
middle term can be performed in either order, so a
parenthesis is not needed; also, we use the usual convention
for raising and lowering indices.

Alternatively, one can decompose the bivector into
\begin{equation}
F = \E +i\B = F^{k0}\bsig_k +\half F^{kj}\bsig_j\bsig_k,\label{2.13b}
\end{equation}
corresponding to the split of an electromagnetic field into``electric
and magnetic parts."

The entire spacetime algebra is obtained from
linear combinations of basis $k$-vectors in (\ref{2.10}).
A generic element $M$ of the STA, called a \textit{multivector},
can thus be written in the \textit{expanded form}
\begin{equation}
M = \alpha + a + F + bi + \beta{}i
   = \sum_{k=0}^4\langle M\rangle_k\,,\label{2.14}
\end{equation}
where $\alpha$ and $\beta$ are scalars, $a$ and
$b$ are vectors, and $F$ is a bivector. This is a
decomposition of $M$ into its $k$-vector parts,
with  $k = 0,1,2,3,4$, where $\langle\ldots \rangle_k$ means
``$k$-\textit{vector part}." Of course,
$\langle M\rangle_0=\alpha$, $\langle M\rangle_1=a$, $\langle M\rangle_2=F$,
$\langle M\rangle_3=bi$, $\langle M\rangle_4=\beta{}i$.

Computations are facilitated by the
operation of \textit{reversion}.
For $M$ in the expanded form (\ref{2.14}) the \textit{reverse}
$\tM$  can be defined by
\begin{equation}
\tM=\alpha + a - F - bi + \beta i\,.\label{2.15}
\end{equation}
Note, in particular, the effect of reversion on the
various $k$-vector parts:

\noindent $\talpha =\alpha,\quad \ta = a,\quad \tF
                 = -F,\quad \tilde i = i\,.$
It is not difficult to prove that
\begin{equation}
(MN)^{\sim} = \tN\tM\,,\label{2.16}
\end{equation}
for arbitrary multivectors $M$ and $N$. For scalar parts
$\langle M\rangle\equiv\langle M\rangle_0$, we have
\begin{equation}
\langle\tM\rangle=\langle M\rangle\qquad\hbox{hence}\qquad\langle MN\rangle=\langle NM\rangle\,.
\label{eq:ScalarPartInSTA}
\end{equation}
The scalar part in STA corresponds to the trace in Dirac's matrix
representation of the algebra.

Besides the inner and outer products defined above, many other products
can be defined in terms of the geometric product. We will need the
\textit{commutator} product $M\times N$, defined for any $M$ and $N$ by
\begin{equation}
  M\times N\equiv \half(MN-NM)=-N\times M\,.\label{2.17}
\end{equation}
We will make frequent use of the ``Jacobi identity" in the form of a
``derivation":
\begin{equation}
  M\times (N\times P) = (M\times N)\times P + N\times (M\times P)\,.
\label{2.18}
\end{equation}
For bivectors $S$ and $F$ we can now expand the geometric product as
follows:
\begin{equation}
SF = S\bdot F + S\times F + S\wedge F\,\label{2.19}
\end{equation}
where $S\times F$ is a bivector and $S\wedge F$ is a pseudoscalar.
Note how this differs from the expansion (\ref{2.7}).

One great advantage of STA is that it provides coordinate-free representations of
both tensors and spinors in the same system. Spinors can be represented as even
multivectors serving as algebraic operators. Using reversion,
it is easy to prove that every even multivector $\psi$ satisfies
\begin{equation}
\psi\tpsi=\rho e^{i\,\beta}.\label{2.20}
\end{equation}
Hence, for $\rho \ne 0$, $\psi$ can be written in the \textit{canonical form}
\begin{equation}
    \psi =(\rho e^{i\,\beta})^\half R,\label{2.21}
\end{equation}
where
 \begin{equation}
    R\tR = \tR R = 1\,.\label{2.22}
\end{equation}
Every such $R$ determines a \textit{Lorentz rotation} of a given multivector $M$:
\begin{equation}
R:   M\quad\rightarrow\quad  M'= RM\tilde{R}\,,\label{2.23}
\end{equation}
and every Lorentz rotation can be expressed in this coordinate-free form.
Construed as an operator in this sense, the quantity $R$ is called a \textit{rotor}
while $\psi$ is called a \textit{spinor}. We shall describe particle kinematics and the
Dirac wave function by spinors in this sense.

The set of all rotors form a multiplicative group called the \textit{rotor group} or
the \textit{spin group} of spacetime.
When $R = R(\tau)$ represents a one parameter family
of rotors (hence of Lorentz rotations), ``angular velocities"
$\Omega$ and  $\Omega'$ are defined by
\begin{equation}
\dot{R} = \half \Omega R = \half R\Omega' \,,\label{2.24}
\end{equation}
or
\begin{equation}
 \Omega = 2\dot{R}\tR =  R\,\Omega'\tR\qquad\hbox{and}\qquad\Omega' =2\tR\dot{R}\,,
\label{2.25}
\end{equation}
where the overdot indicates derivative. It follows from (\ref{2.22}) and
(\ref{2.14}) that $\Omega$ and $\Omega'$ are bivectors.

We represent each spacetime point as a vector $x=x^\mu \gamma_\mu$
with rectangular coordinates $x^\mu=\gamma^\mu\cdot x$. The \textit{vector derivative}
with respect to the point $x$ is defined by $\nabla=\partial_x =\gamma^\mu
\partial_{x^\mu}$. As $\nabla$ is a vectorial operator, we can use (\ref{2.7}) to
decompose the derivative of a $k$-vector field $K=K(x)$ into
\begin{equation}
\nabla K=\nabla\bdot K + \nabla\wedge K\,,\label{2.26}
\end{equation}
where the terms on the right can be identified, respectively, with the usual \textit{divergence} and \textit{curl} in tensor calculus.

Besides the STA definitions and relations given above, many others can be
found in the references. We shall introduce some of them as needed.

\section{Classical Particles with Spin}\label{sec:III}

Classical models of the electron as a point particle
with spin were first formulated by Frenkel \cite{Frenkel} and
Thomas \cite{Thomas}, improved by Mathisson \cite{Mathisson},
and given a definitive form by Weyssenhoff \cite{Weys47}.
They have been revisited from time to time by many investigators \cite{Corben,
Gursey,Rivas01}, including others to be cited below.
They are of interest mainly for the insight they bring to the interpretation of
quantum mechanics. But they are also of practical value, for example,
in the study of spin precession \cite{BMT,Hest03b}
and tunneling \cite{Beauregard,DLGSC,Rivas98}.

In Weyssenhoff's analysis the models fall into distinct classes,
differentiated by the assumption that the electron's spacetime history
is timelike in one and lightlike in the other. The timelike case has
been studied by many investigators, such as those cited above. Ironically,
the lightlike case, which Weyssenhoff regarded as far more interesting, has been
generally ignored. Without being aware of his analysis,
I arrived at similar conclusions from my study of real Dirac theory.
This paper revisits Weyssenhoff theory with new mathematical tools to simplify and
extend it.

Our first task is to reformulate Weyssenhoff theory in the language of
spacetime algebra (STA) and show how it is simplified,
clarified and extended. A decided advantage of STA is its uniform use of spinors in
both classical and quantum theory. For one thing, this makes it easier to
relate classical models to solutions of the Dirac equation \cite{Hest03b}.
Although we are most interested in the lightlike case, we treat the timelike
case in enough detail to compare the two cases. Besides, the timelike models
can be used for other particles with spin besides electrons, including atoms.

We consider a particle with spacetime \textit{history} $z = z(\tau)$
and \textit{proper velocity} \cite{Hest03b}
\begin{equation}
u \equiv \dot{z} = \frac{dz}{d\tau}\,.\label{3.1}
\end{equation}
For the time being, we allow $u$ to be either timelike, with $u^2=1$, or lightlike,
with $u^2=0$, so in either case
\begin{equation}
\dot{u}\bdot u = 0\,.\label{3.2}
\end{equation}
In the timelike case, of course, the parameter $\tau$ is taken to be
 proper time, but in the lightlike case it must be regarded as arbitrary for the
time being.

Suppose the particle has \textit{proper momentum} $p = p(\tau)$ and intrinsic angular
momentum (\textit{spin}) represented by a bivector $S = S(\tau)$. Equations of motion
are obtained from general conservation laws for momentum and angular momentum.
However, to formulate these laws correctly it is essential to note that
just as $p$ contributes an orbital momentum, $S$ contributes an intrinsic
part to the momentum, so we cannot make the usual assumption $p = mu$. Instead,
the relation between $p$ and $u$ depends on the dynamics of motion. The same can
be said about the relation between $S$ and $u$, although we shall assume
that it is restricted by the \textit{kinematical constraint}
\begin{equation}
   S\bdot u = 0\,.\label{3.3}
\end{equation}
This reduces the degrees of freedom in $S$ to three at most.

The noncollinearity of $p$ and $u$ raises a question about how
mass should be defined. Without prejudicing the issue, it is
convenient to introduce a \textit{dynamical mass} defined by
\begin{equation}
  m \equiv p\bdot u\,.\label{3.3b}
\end{equation}
This quantity is well defined for both timelike and lightlike particles, so lightlike
particles can have mass in this sense.
However, the value of this mass obviously depends on the choice of time parameter $\tau$,
which, so far, is not well defined in the lightlike case. This is our first hint that mass
is intimately related to intrinsic time scaling of electron histories.

Without presupposing a relation between momentum and velocity,
\textit{momentum conservation} can be given the general form
\begin{equation}
\dot{p} = f\,,\label{3.4}
\end{equation}
where the \textit{proper force} $f$ describes transfer of momentum through
external interactions.

\textit{Angular momentum conservation} is governed by
\begin{equation}
\dot{S} = u\wedge p + \Gamma\,,\label{3.5}
\end{equation}
where the \textit{proper torque} $\Gamma$ describes angular momentum transfer
through external interactions. To see where (\ref{3.5}) comes
from, we introduce the \textit{orbital angular momentum} $p\wedge z$ so the
\textit{total angular momentum} is
\begin{equation}
  J \equiv p\wedge z + S\,.\label{3.6}
\end{equation}
The angular momentum conservation then has the more familiar form
\begin{equation}
  \dot{J} = f\wedge z + \Gamma\,,\label{3.7}
\end{equation}
where $f\wedge z$ is the orbital torque. The equivalence of (\ref{3.7})
with (\ref{3.5}) follows immediately by differentiating (\ref{3.6}).

Equation (\ref{3.7}) tells us that $J$ is a constant of
motion for a free particle with spin. Spin is not separately conserved. According
to (\ref{3.6}), angular momentum can be exchanged
back and forth between orbital and spin parts.

To get well defined equations of motion from the conservation laws we need to
specify the interactions.  For a particle with charge $q$ and dipole moment
$M = M(\tau)$, we use
\begin{equation}
  f = q F\bdot u + \nabla F \bdot M\label{3.8}
\end{equation}
  where $F = F(z)$ is an applied electromagnetic (bivector) field [aptly called
the \textit{Faraday}] and the vector derivative $\nabla=\gamma^\mu \partial_\mu$ operates
on $F$. The first term on the right of (\ref{3.8}) is the standard ``Lorentz force,"
while the second is a force of ``Stern-Gerlach" type. The latter is
accompanied by the torque
\begin{equation}
  \Gamma = F\times M\,.\label{3.9}
\end{equation}
The ``interaction laws" (\ref{3.8}) and
(\ref{3.9}) must be supplemented by a ``constitutive equation"
expressing $M$ as a function of $S$ and $u$. In general $M$ can include both
electric and magnetic dipole moments. The general
constitutive constraint
\begin{equation}
  M\times S = 0 \label{3.10}
\end{equation}
is satisfied by the standard constitutive equation $M = \kappa S$ for a
magnetic moment, but also by the more general relation
\begin{equation}
   M = \kappa S e^{i\beta}\,.\label{3.11)}
\end{equation}
The factor $e^{i\beta}$ is a ``duality rotation" relating a magnetic moment
$i S\kappa \sin\,\varphi$ to an electric moment
$S\kappa \cos\,\varphi$. As noted later, the magnetic moment (density) in Dirac theory has
this form.  In the lightlike case we shall see that there is an alternative way to satisfy
(\ref{3.10}).

From the specific form of the interaction laws we can prove that the magnitude of the
spin is a conserved quantity. First note that (\ref{3.3}) implies
\begin{equation}
  S \wedge S = 0\qquad\hbox{so}\qquad S^2 = S\bdot S\,.\label{3.12}
\end{equation}
This follows from the identity
$$  (S \wedge S) u = (S \wedge S)\bdot u = 2 S\wedge (S\bdot u)\,.$$
From (\ref{3.5}) with (\ref{3.9}) we get
$$
 \half\frac{d}{d\tau}S^2 = S\bdot\dot{S}
    = S\bdot( u\wedge p) + S\bdot(M\times F)\,.
$$
The first term on the right vanishes because
$$  S\bdot( u\wedge p) = (S\bdot u)\bdot p = 0\,,$$
and the second term vanishes by (\ref{3.10}) and the identity
\begin{equation}
 S\bdot( F\times M) = \langle SFM\rangle = F\bdot (M\times S)\,.\label{3.13}
\end{equation}
Therefore
\begin{equation}
  S^2 = -|S|^2 = \hbox{constant}\,. \label{3.14}
\end{equation}
The negative sign appears because (\ref{3.3}) implies that $S$
cannot be a timelike bivector.

Up to this point, our equations apply equally to timelike and lightlike particles.
To go further we must consider each of these cases separately.
There are three distinct possibilities which have been studied in the past.
Most investigators have been attracted by the possibility of choosing
$|S|=\hbar/2$ to make contact with Dirac theory. The possibility of coupling this
with a timelike velocity has been most thoroughly studied by Corben \cite{Corben}.
However, his model does not recognize the intimate connection between electron spin
and phase. The alternative possibility of coupling with a lightlike velocity
overcomes that deficiency. I confess to struggling with that model for many years
before I realized that for a lightlike particle the spin must be a lightlike
bivector. Shortly thereafter I learned that Weyssenhoff had figured that out long
before \cite{Weys47}.

There are many more things to say about the timelike case, but we shall concentrate
on the more interesting lightlike case of Weyssenhoff. The tools of STA enable us to
develop this case much farther than he could, indeed, to a fairly definitive and
satisfactory conclusion.

\section{Free Particle Motion}\label{sec:IV}

We gain crucial insight into the ``kinematics of spin" from the free particle
solution. The solution has the same form for both timelike and lightlike
cases, though we shall see that there are significant differences in physical
interpretation.

For a free particle it follows from (\ref{3.4}), (\ref{3.7}) and (\ref{3.6})
that $p$ and $J = p\wedge z + S$ are constants of motion.
We seek a general solution with constant mass  $m_e=p\bdot u=p\bdot \dot{z}$.
Integrating this relation, we obtain
\begin{equation}
  p\bdot(z-z_0) = m_e\tau\,,\label{5.1}
\end{equation}
where $z_0$ is the particle position at $\tau = 0$. Angular momentum conservation
gives us
\begin{equation}
  (z-z_0)\wedge p = S(\tau) - S_0\,,\label{5.2}
\end{equation}
where $S_0$ is the initial value of the spin. Adding these equations and multiplying by
$p^{-1}=p/p^{2}$ we get an equation for the \textit{particle history}:
\begin{equation}
  z = [S(\tau) - S_0]\,p^{-1} + m_e p^{-1}\tau + z_0\,. \label{5.3}
\end{equation}
We can write this in the form
\begin{equation}
  z(\tau) = r(\tau) + x(\tau)\,,\label{5.4}
\end{equation}
where a center of motion is defined by
\begin{equation}
  x = \frac{m_e}{p}\,\tau + z_0 - S_0\bdot p^{-1}\,,\label{5.5}
\end{equation}
and a radius vector is defined by
\begin{equation}
 r = S(\tau)\bdot p^{-1}\,.\label{5.6}
\end{equation}
Differentiating, we obtain
\begin{equation}
 \dot{r} = \dot{S}\bdot p^{-1}=(u\wedge p)\bdot p^{-1}=u-m_ep^{-1}\,.\label{5.7}
\end{equation}
It follows that $\dot{r} \bdot r=0$, so $r^2$ is a constant of motion, and
$r = r(\tau)$ must be a rotating vector.
Hence (\ref{5.4}) describes a helix
in spacetime centered on the timelike straight line (\ref{5.5}).

We get an explicit form for the temporal behavior of $S$ and $r$ by
introducing a rotor $R = R(\tau)$  so that
\begin{equation}
S = R S_0 \tR\,.\label{5.8}
\end{equation}
Then (\ref{5.6}) is consistent with
\begin{equation}
  r = R r_0 \tR \label{5.9}
\end{equation}
provided
\begin{equation}
  p\bdot\Omega = 0\label{5.10}
\end{equation}
for $\Omega = 2\dot{R}\tR$. Furthermore
\begin{equation}
  \dot{z} = u = R u_0\tR \label{5.11}
\end{equation}
if $\Omega$ is a constant, in which case $R'$ integrates to
\begin{equation}
 R = e^{\smallhalf\Omega\tau}\,.\label{5.12}
\end{equation}
From this we conclude that all aspects of the motion are determined by the
scalar, vector and bivector constants of motion, $m_e$, $p$ and $\Omega$. The values of
these constants are left to be determined by other considerations.

These results apply to both timelike and lightlike cases. In the timelike case, if one
assumes $|S|=\hbar/2$, it is easy to prove that
\begin{equation}
  |\Omega|=\omega_e \equiv\frac{2m_e c^2}{\hbar}  \approx 1.5527\times10^{21}
\hbox{rad/sec},\label{5.13}
\end{equation}
which is the \textit{zitter frequency} of Schroedinger where $m_e$ is the electron
rest mass.

Moreover, writing $p=m_e v$, we can express the proper time in the form
$\tau=v\bdot z=v\bdot x$, so (\ref{5.12}) can be expressed in the form
\begin{equation}
 R = e^{\smallhalf\i\, \omega_e \tau}=e^{\i\, p\bdot x/\hbar}\,,\label{5.13a}
\end{equation}
which, with $\i$ as a unit spacelike bivector, has been shown elsewhere \cite{Hest03b} to
correspond precisely to plane wave solutions of the Dirac equation.

The same solution applies to the lightlike case, but with the additional constraint
$\omega_e \lambda_e=c$, so the \textit{electron zitter radius} has the fixed value
\begin{equation}
  \lambda_e \equiv\frac{\hbar}{2m_e c}  \approx 1.9308\times10^{-3}\,
{\AA}.\label{5.13b}
\end{equation}
Regarded as a solution of the Dirac equation, the spinor (\ref{5.13a}) describes a
congruence of either timelike or lightlike histories, depending on the choice of velocity
observable.
Our next task is to generalize these insights to include interactions.

\section{Lagrangian for a lightlike particle with spin}\label{sec:V}

In Section \ref{sec:III} we postulated equations of motion and a number of
relations among velocity, spin, momentum and mass variables. To ensure that all this
constitutes a complete, coherent and self-consistent dynamical system we show
that it can be derived from a single Lagrangian, but with introduction of further
constraints.

Conventional Lagrangians cannot be used for lightlike
particles, owing to the absence of proper time as a natural parameter.
Weyssenhoff \cite{Weys51} addressed this problem by deriving
parameter invariant Lagrangians for both timelike and lightlike particles.
However, he encountered problems that limited his treatment to the free particle
case.
Kr\"uger \cite{Kruger97} resolved the problem for a lightlike particle by
introducing a Lagrangian defined in terms of particle path curvature (i.e.
acceleration) instead of velocity.
However, his model does not have all the features that we are looking for, so we
employ a different approach here.

We take full advantage of STA by using it to construct a spinor-based Lagrangian.
Proca \cite{Proca} was the first to use Dirac spinors for describing classical
particles. Barut and Zanghi \cite{Barut&Zanghi} used the same approach to model
electron zitter. Gull \cite{Gull} translated their model into STA and noted that it
lacked a magnetic moment.

Here we see that spinors are especially helpful for building lightlike constraints
into the Lagrangian. The resulting spinor equations provide a superior dynamical
model for a lightlike particle with spin, mass and zitter, which, as we see later, is
intimately related to the Dirac equation.
Thus, we are able to complete Weyssenhoff's program
to construct dynamical equations for a lightlike particle with zitter and
electromagnetic interactions.

Continuous ``motion" of a particle in spacetime is represented by a curve
$z=z(\varphi)$ and its derivative $z'=dz/d\varphi$, where $\varphi$ is an
affine parameter for which the physical interpretation is initially unspecified.
The kinematic structure of this curve is described by a spinor
$\psi=\psi(\varphi)$ and its derivative $\psi'$.

The dynamics of motion in an external electromagnetic field $F(z)=\nabla\wedge A$
with vector potential $A=A(z)$ is determined by a Lagrangian
$L=L(z,\psi,P)$ of the form
\begin{multline}
 L=\left<-\hbar\psi'\gamma_+\gamma_1\tpsi+P(z'-\psi\gamma_+\tpsi)\right.\\ \left.
+qA\psi\gamma_+\tpsi+q\lambda_e F\psi\gamma_+\gamma_1\tpsi\right>, \label{4.1}
\end{multline}
with $\gamma_\pm \equiv \gamma_0 \pm \gamma_2$ and units: $c=1, \hbar=m_c=$ fixed reference
mass. The coupling constants are charge $q$ and a length $\lambda_e$, which we anticipate
identifying as the free electron zitter radius (\ref{5.13b}), so it amounts to
introducing the electron rest mass $m_e$ as a coupling constant. A vectorial Lagrange
multiplier $P$ relates the two kinds of kinematic variable.

The \textit{method of multivector differentiation} is the simplest and most elegant way to
derive equations of motion \cite{Lasenby93}.
For a Lagrangian that is homogeneous of degree one in derivatives, variation with
respect to a multivector variable
$X=X(\varphi)$ yields the multivector Lagrange equation
\begin{equation}
\delta_XL\equiv \partial_X L -\partial_\varphi(\partial_{X'}L)=0,\label{4.1b}
\end{equation}
where $X'=\partial_\varphi X$.

Variation of the Lagrangian with respect to $P$ obviously gives
\begin{equation}
  z'=\psi\gamma_+\tpsi\equiv w\,,\label{4.2}
\end{equation}
which defines a \textit{particle velocity}  $w$ in terms of the spinor $\psi$.

The result of varying position
vector $z$ is the \textit{force law}
\begin{equation}
  p'=qF\cdot w+\nabla F\cdot M_w,\label{4.3}
\end{equation}
with \textit{momentum vector} $p$ defined by the canonical expression
\begin{equation}
p\equiv P-qA,\label{4.4}
\end{equation}
and electromagnetic moment bivector defined by
\begin{equation}
 M_w\equiv q\lambda_e \psi\gamma_+\gamma_1\tpsi.\label{4.5}
\end{equation}
Of course, the vector derivative $\nabla$ in (\ref{4.3}) operates only on $F$ and not
on $M$.

The result of varying spinor $\psi$ in the Lagrangian is the dynamical spinor equation
\begin{equation}
 \hbar\psi'\gamma_+\gamma_1=-p\psi\gamma_+
+q\lambda_e F\psi\gamma_+\gamma_1\,. \label{4.6}
\end{equation}
From this one can easily derive the equation of motion
\begin{equation}
 S_w'=w\wedge p+F\times M_w \label{4.7}
\end{equation}
for a \textit{spin bivector} defined by
\begin{equation}
S_w\equiv \frac{\hbar}{2}\psi\gamma_+\gamma_1\tpsi. \label{4.8}
\end{equation}
One can also derive the equation
\begin{equation}
\frac{\hbar}{2}w'=\frac{2}{\hbar}p\cdot S_w+q\lambda_e F\cdot w, \label{4.9}
\end{equation}
which relates particle acceleration to spin and momentum.

The Lagrangian has delivered a system of coupled equations for particle velocity, spin
and momentum. Some simplifications are still needed to facilitate physical
interpretation and analysis.

\section{Observables and Dynamical Structure}\label{sec:VI}

Consider implications of the canonical decomposition $\psi=(\rho
e^{i\,\beta})^\half R$. The rotor $R=R(\varphi)$ determines a one parameter family of
Lorentz rotations
\begin{equation}
e_\mu=e_\mu(\varphi)=R\gamma_\mu\tR \label{4.10}
\end{equation}
that transforms a fixed orthonormal frame of vectors into an intrinsic \textit{comoving frame} following the particle. The comoving frame is coupled to the particle
velocity by
\begin{equation}
 w\equiv z'=\psi\gamma_+\tpsi =\rho u\,,\label{4.11}
\end{equation}
where a new rescaled velocity has been defined by
\begin{equation}
u\equiv R\gamma_+\tR=e_0+e_2. \label{4.12}
\end{equation}
Thus, we identify $\rho$ as a \textit{time scale parameter!}

The \textit{duality factor} $ e^{i\,\beta}$ rotates multivectors into their duals,
as shown by the equation
\begin{equation}
e^{i\,\beta}\gamma_\mu=\gamma_\mu \cos\beta +i\gamma_\mu \sin\beta
=\gamma_\mu e^{-i\,\beta}.\label{4.13}
\end{equation}
Note that anticommutivity of vectors with the pseudocalar has eliminated the duality
factor from (\ref{4.11}). However, the duality factor commutes with bivectors.
So for the spin bivector we find
\begin{eqnarray}
S_w&=& \frac{\hbar}{2}\psi\gamma_+\gamma_1\tpsi
=\frac{\hbar}{2}\rho R\gamma_+ \tR R\gamma_1\tR e^{i\,\beta}\nonumber\\
&=&\frac{\hbar}{2}\rho u e_1(e^{i\beta}).  \label{4.14}
\end{eqnarray}
This makes the relation of spin to velocity $u=e_0+e_2$ explicit. Further simplification
is possible by eliminating the duality factor.

Note that the unit pseudoscalar can be written
$i=\gamma_0\gamma_1\gamma_2\gamma_3= e_0e_1e_2e_3 = e_0e_2e_3e_1$, so
$ue_0e_2 =u$ implies $iu=-ui = ue_1e_3$. Consequently, the effect of the duality factor
multiplying a null vector is equivalent to a rotation, and (\ref{4.14}) is proportional
to the form
\begin{eqnarray}
e^{i\,\beta}ue_1&=&ue^{-i\,\beta}e_1=ue^{i e_0 e_2 \beta}e_1
=ue^{e_3e_1\beta}e_1\nonumber\\
&=&u(e_1\cos\beta+e_3\sin\beta).
\label{4.15}
\end{eqnarray}
In other words, the net effect of $e^{i\,\beta}$ in (\ref{4.14}) is to rotate $e_1$
into $e_3$.
Since the duality factor appears in every term of the spinor equation (\ref{4.6}),
it can be eliminated by absorbing it into the rotor, as shown by
\begin{equation*}
\hspace{-.6cm}\psi\gamma_+=\rho^\half Re^{i\,\beta/2}\gamma_+
=\rho^\half (Re^{\gamma_1\gamma_3\beta/2})\gamma_+,
\end{equation*}
so that (\ref{4.14}) reduces to
\begin{equation}
S_w= \frac{\hbar}{2}\psi\gamma_+\gamma_1\tpsi
=\frac{\hbar}{2}\rho ue_1=\rho S,  \label{4.15a}
\end{equation}
where we have introduced a \textit{rescaled spin bivector} $S$. Note  that the spin has
a dual form
\begin{equation*}
\hspace{-3.2cm}-iS=\frac{\hbar}{2}e_3e_1ue_1=\frac{\hbar}{2}e_3u,
\end{equation*}
Thus, we can characterize the \textit{spin bivector} by the compact equations
\begin{equation}
S\equiv \frac{\hbar}{2} ue_1=isu,\qquad \hbox{where} \qquad s\equiv\frac{\hbar}{2} e_3
\label{4.15b}
\end{equation}
defines a \textit{spin vector}.

\subsection{Momentum, mass and velocity}

Particle mass relates momentum to velocity. When momentum is not collinear with
velocity there are two distinct concepts of mass. First, assuming that the kinetic
momentum $p$ points to the future,  we can define a positive  mass $m_p$ in the
usual way: $p^2 =m_p^2>0$.
Second, projection of momentum $p$ onto the particle velocity $u$ defines a \textit{dynamical mass} $m\equiv p\bdot u$.
This turns out to be the most significant mass variable.

To specify the relation of momentum to velocity more completely, we note that
the comoving frame $\{e_\mu\}$ is not uniquely determined by its relation to
the particle velocity in (\ref{4.12}) and spin in (\ref{4.15b}). Consequently, we are
free to constrain $e_0$ to lie in the $p\wedge u$ plane, as specified by the
equation
\begin{equation}
p\wedge u\wedge e_0=p\wedge e_2\wedge e_0=0. \label{4.16}
\end{equation}
Hence, we can introduce mass
parameters $m_1$ and $m_2$ so that
\begin{eqnarray}
\hspace{-.4cm}&&p =m_1e_0 -m_2e_2, \nonumber\\
\hspace{-.4cm}&&\hbox{with} \qquad  m\equiv p\bdot u=m_1+m_2. \label{4.17}
\end{eqnarray}
The various masses are thus related by
\begin{equation}
\frac{m_p^2}{m^2}=\frac{p^2}{(p\bdot u)^2}=\frac{m_1^2-m_2^2}{(m_1+m_2)^2}
=\frac{m_1-m_2}{m_1+m_2}.
\label{4.18}
\end{equation}
Functional values for the masses will be determined by the equations of motion.

Now the relation of momentum to particle velocity can be expressed in the simple form
\begin{equation}
up =m(1+ e_2e_0) \quad \hbox{with} \quad u\wedge p =me_2e_0.\label{4.19}
\end{equation}
Likewise, it follows from (\ref{4.17}) that $p\bdot e_1$ and $p\bdot e_3=p\bdot s=0.$
Hence, the relation of momentum to spin is reduced to the simple equations
\begin{equation}
p\bdot S=\frac{\hbar}{2}p\bdot (ue_1)=\frac{\hbar}{2}me_1,\label{4.20}
\end{equation}
\begin{eqnarray}
p\wedge S=p\wedge (isu)=ip\cdot(us)= ims \nonumber\\
\hbox{or} \quad
ms=-i(p\wedge S)=p\cdot(iS).\label{4.21}
\end{eqnarray}
These relations considerably simplify the velocity
equation of motion (\ref{4.9}) and, as we see later, the spinor equation (\ref{4.6}) by
reducing direct coupling with momentum to the scalar variable $m=p\bdot u$. One more
simplification is needed.

\subsection{Time scaling and gauge transformations}

For purposes of measurement, we must relate the affine parameter $\varphi$ to some observable
time scale.
This can be done by projecting the electron's null velocity onto a \textit{timelike reference
history}
$x=x(\tau)$ with \textit{proper time} $\tau$
and \textit{ velocity}
\begin{equation}
v\equiv\dot x=dx/d\tau,\label{4.21b}
\end{equation}
thus associating a proper time with the electron motion. The electron velocity
(\ref{4.2}) is thus rescaled to proper time by
\begin{equation}
 \rho \equiv v\bdot w=\frac{d\tau}{d\varphi},\quad\hbox{so}\quad
w=\frac{d\tau}{d\varphi}\frac{dz}{d\tau}\equiv \rho u.\label{4.22}
\end{equation}
Accordingly, the rescaled  particle velocity is given by
\begin{equation}
\frac{dz}{d\tau}\equiv \dot{z}=u=e_0+e_2=R\gamma_+\tR. \label{4.23}
\end{equation}
This leaves us with the problem of specifying the reference history and determining the time
scale factor $\rho$. We have considerable leeway in choosing the reference history, the only requirement being
that its relation to the electron history $z(\tau)$ be well defined. Two obvious choices
for reference velocity are $v=p/m_p$ or $v=e_0$, but we shall see that they have drawbacks,
so it is best to leave the option open until a decision is needed.

Recall that $\varphi$ was originally introduced as an arbitrary affine parameter in the
Lagrangian (\ref{4.1}). In fact, the Lagrangian time scale depends on the spinor gauge, as
we now prove. Dropping the prime notation for $d/d\varphi$ from here on, and using the prime to indicate a  change of variable, we note that a time
scale gauge transformation of the spinor can be defined as a \textit{boost}
\begin{equation}
 \psi \qquad \rightarrow\qquad
\psi'=\psi e^{\half \alpha \gamma_2\gamma_0}\label{4.24}
\end{equation}
which induces a scale change
\begin{equation}
 \psi\gamma_+\tpsi \qquad \rightarrow\qquad
\psi'\gamma_+\tpsi'= e^ \alpha \psi\gamma_+\tpsi.\label{4.25}
\end{equation}
Combining this with the change of variable
\begin{equation}
 \frac{d}{d\varphi} \qquad \rightarrow\qquad
\frac{d}{d\varphi'}=\frac{d\varphi}{d\varphi'}=e^\alpha\frac{d}{d\varphi},\label{4.26}
\end{equation}
we have a \textit{time scale gauge transformation} that induces a change of the Lagrangian
(\ref{4.1}) to
\begin{eqnarray*}
 L'& = &\left<-\hbar
e^\alpha\frac{d}{d\varphi}(e^{\alpha/2}\psi)\gamma_+\gamma_1\tpsi e^{\alpha/2}\right>\\
& + &\left< P(e^\alpha\frac{dz}{d\varphi}-e^\alpha\psi\gamma_+\tpsi)
+qAe^\alpha\psi\gamma_+\tpsi\right.\nonumber\\ &&\left.
\hspace{1.6cm} +\ q \lambda_e Fe^\alpha\psi\gamma_+\gamma_1\tpsi \right>.
\end{eqnarray*}
Since the first term transforms differently from the other terms, the Lagrangian is not time
scale invariant. Consequently, the time parameter must have a physical significance, though
that was not imposed in writing the Lagrangian.

We can fix the time scale gauge by requiring $\rho=d\tau/d\varphi=\pm 1$, so proper time
becomes our intrinsic time variable. That simplifies our model considerably. The choice is
most conveniently expressed by adopting the constraint
\begin{equation}
\psi\tpsi=\rho e^{i\beta}=\pm 1, \label{4.27}
\end{equation}
which suppresses the superfluous $ \beta$ variable as well. There are reasons to associate
the minus sign with an antiparticle (positron), but we will not explore that possibility
here and will stick with the plus sign for the rest of this paper.

We could have imposed the constraint (\ref{4.27}) at the beginning by replacing the spinor
$\psi$ with the rotor $R$ in the Lagrangian. That was not done for two good reasons.
First, it avoids messing with the constraint $R\tR=1$, which complicates variational
derivatives of the Lagrangian. Second, the Dirac wave function also has a factor $\rho
e^{i\beta}$, so we want to understand how that relates to the similar factor here.  Indeed,
if the Dirac equation is derivable from a superposition of particle paths, as might be the
case, then the individual proper times must be replaced by a common time variable, so $\rho$
should emerge as a common time scale factor for the ensemble of paths.

To ascertain the significance of our gauge choice, we consider a class of gauge
transformations with the form
\begin{equation}
 \psi \qquad \rightarrow\qquad
\psi'=\psi e^{\Gamma\, \chi},\label{4.28}
\end{equation}
where, as in (\ref{4.24}), the generator $\Gamma$ is a constant bivector constructed from the
vectors $\gamma_\mu$. For $\Gamma=\gamma_2\gamma_1$, this is the form for an electromagnetic
gauge transformation in the Dirac equation when it is accompanied by the transformation of
vector potential:
\begin{equation}
\frac{q}{\hbar}A \qquad \rightarrow\qquad
\frac{q}{\hbar}A'=\frac{q}{\hbar}A+\nabla \chi.\label{4.29}
\end{equation}
Inserting this into the Lagrangian (\ref{4.1}), we find gauge
invariance if the following two terms cancel:
\begin{equation}
(\nabla\chi)\bdot (\psi'\gamma_+\tpsi')= w'\bdot \nabla \chi, \label{4.30}
\end{equation}
\begin{eqnarray}
\left\langle\frac{d\psi'}{d\varphi}\gamma_+\gamma_1 \tpsi'\right\rangle&=&\frac{d\chi}{d\varphi}
\left\langle\psi'\gamma_2\gamma_1\gamma_+\gamma_1\tpsi'\right\rangle\nonumber\\
&=&-\frac{d\chi}{d\varphi}
\left\langle\psi'\tpsi'\right\rangle. \label{4.31}
\end{eqnarray}
Cancelation occurs if $\left\langle\psi'\tpsi'\right\rangle=1$, which is precisely our constraint
(\ref{4.27}).

This is an intriguing if not ironic result. Recall that Weyl originally introduced gauge
transformations as an extension of General Relativity to incorporate length scale
invariance. Subsequently, he renounced that idea and introduced the notion of gauge
invariance that has become standard in quantum mechanics. Here we have come full circle
to find electromagnetic gauge invariance associated with time scale invariance. Perhaps Weyl
had the right idea in the first place.

\subsection{Proper equations of motion and first integrals}

Now we are prepared to reformulate our equations of motion in their simplest form, with
proper time as the independent variable.  Using the simplification (\ref{4.20}), we can
recast the equation (\ref{4.9}) for particle velocity in the rescaled form
\begin{equation}
\dot{u}=\ddot{z}=\omega e_1+\frac{q}{m_e}F\cdot u-\dot{\rho} u, \label{4.32}
\end{equation}
where a  \textit{dynamical
zitter frequency} has been defined by $\omega\equiv 2m/\hbar$.

The momentum equation (\ref{4.3}) can be written
\begin{equation}
 \dot{p} =qF\cdot u+\nabla\Phi,\label{4.33}
\end{equation}
where the \textit{spin potential} for the gradient force is given by several equivalent forms:
\begin{eqnarray}
\Phi=\Phi(\tau,z)&\equiv& \frac{q}{m_e} S(\tau)\bdot F(z)\nonumber\\
&=&\frac{q}{m_e} (isu)\bdot F=q\lambda_e F\bdot(ue_1).\label{4.34}
\end{eqnarray}
For future reference, we note that
\begin{eqnarray}
&&\hspace{-.7cm}\frac{m_e}{q} \dot{\Phi}=\frac{m_e}{q}(\partial_\tau +u\bdot\nabla)\Phi\nonumber\\
&=&(\partial_\tau S)\bdot F+S\bdot(\dot{z}\bdot\nabla F)
=\dot{S}\bdot F+S\bdot \dot{F}.\label{4.35}
\end{eqnarray}
Finally, from (\ref{4.7}) we get the rescaled equation of motion for spin
\begin{equation}
\dot{S}=u\wedge p+\frac{q}{m_e} F\times S-\dot{\rho} S,\label{4.36}
\end{equation}
wherein the coupling with momentum can be simplified by (\ref{4.19}) to
 $u\wedge p=me_2e_0$.

From the three \textit{proper equations of motion} (\ref{4.32}), (\ref{4.33}) and (\ref{4.36}),
we easily derive
\begin{eqnarray}
\frac{d(\rho m)}{d\tau}&=&\frac{d( p\bdot w)}{d\tau}
=\dot{p}\bdot w+p\bdot \dot{w}\nonumber\\
=\frac{q}{m_e}&&\hspace{-.7cm}\frac{d(\rho S)}{d\tau}\bdot F
+\rho u\bdot\nabla\Phi
 = \frac{d}{d\tau}(\rho \Phi), \label{4.37}
\end{eqnarray}
where the identity (\ref{3.13}) has been used with (\ref{4.35}).
Thus we obtain an integral of motion for the  electron mass:
\begin{equation}
\rho m=m_e +\rho\Phi. \label{4.38}
\end{equation}
The integration constant is readily identified as the free electron mass $m_e$, and we
recognize $\Phi$, with its several different forms (\ref{4.34}), as a variable \textit{mass
shift} due to interaction.

To learn more about the relation between mass and momentum, we note that the common
interaction term in eqs. (\ref{4.32}) and (\ref{4.33}) can be eliminated to get
\begin{equation}
 \frac{d}{d\tau}(p-m_e\rho u) =\nabla\Phi-m\omega_e e_1.\label{4.39}
\end{equation}
This suggests that we define a \textit{relative momentum vector}
\begin{equation}
\pi\equiv p-m_e\rho u.\label{4.40}
\end{equation}
It then follows that \begin{equation}
\pi\bdot \dot{\pi}=\frac{d}{d\tau}\left(\frac{\pi^2}{2}\right)=\pi\bdot \nabla
\Phi.\label{4.41}
\end{equation}
This describes the rate that spin interaction energy is converted to mass. It vanishes when
the impressed field $F$ is constant, in which case $\pi^2$ is a constant of motion.

Since $m= p\bdot u=m_1 +m_2$, the various mass parameters are related by
\begin{equation}
\pi^2= p^2-2m_em\rho=m(m_1-m_2-2m_e\rho).\label{4.42}
\end{equation}
Thus, we have two equations of change (\ref{4.37}) and (\ref{4.41}) for three mass parameters
$\{m_1, m_2, \rho\}$ or $\{m, m_1- m_2, \rho\}$
By inspection, no other relations among these parameters can be derived from the dynamical
equations. This leaves us with a gauge freedom to fix one relation among them. There are two
possibilities with clear physical meaning. On the one hand, we could require
$m_1^2-m_2^2=p^2=(p\bdot u)^2=m^2$, so there is only one mass parameter.
\textit{Alternatively, we can require $\rho=1$.
This appears to be the better choice, as it simplifies the equations of motion and,
as we have already noted, it accommodates gauge invariance.} Here we have learned how it
relates to mass parameters.
However, before committing to a gauge choice, there is more to say about the equations of
motion.

\subsection{Rotation dynamics}

So far we have established a system of three equations of
motion, (\ref{4.32}), (\ref{4.33}) and (\ref{4.36}), for electron velocity, momentum and
spin. Now it will be shown how these three equations can be reduced to a single spinor
equation. The reduction is important for two reasons: First, it simplifies solving
the equations of motion. Second, it greatly facilitates comparison with Dirac theory.

Reduction is made possible by expressing the momentum in the form
\begin{eqnarray}
\hspace{-.4cm}&&p=m_1e_0-m_2e_2=Rp_0\tR, \nonumber\\
\hspace{-.4cm}&&\hbox{where}\quad p_0\equiv m_1\gamma_0-m_2\gamma_2.\label{4.43}
\end{eqnarray}
Since the velocity is given by $u=e_0+e_2=R\gamma_+\tR$ and the spin is given by
$S=(\hbar/2)ue_1=(\hbar/2)R\gamma_+\gamma_1\tR$, the evolution of all three observables is
determined by
the rotation dynamics of the comoving frame $e_\mu=R\gamma_\mu \tR$.
That, in turn, is determined by a single equation of motion for the rotor $R=R(\tau)$.
The normalization condition $R\tR=1$ implies that the rotor evolution
equation has the general form
\begin{equation}
\dot{R}=\frac{1}{2}\Omega R,\label{4.44}
\end{equation}
where the bivector \textit{rotational velocity} $\Omega=\Omega(\tau)$ is a
specified function. It follows immediately that equations of motion for the $e_\mu$
are given by
\begin{equation}
\dot{e}_\mu=\Omega \bdot e_\mu.\label{4.45}
\end{equation}
Our problem is therefore to determine the functional form of $\Omega$.

Of course, the equations of motion for velocity and spin were initially derived from the
spinor equation (\ref{4.6}). That equation is difficult to handle as it stands; however, we
are now prepared to put it in more tractable form. Excluding the duality factor for reasons
already explained, we can write the spinor in the form $\psi=\rho^{1/2} R$, so (\ref{4.43})
gives us
\begin{equation}
p\psi=\psi p_0.\label{4.46}
\end{equation}
Using the identity $\gamma_0\gamma_+=-\gamma_2\gamma_+$ we have
\begin{eqnarray}
\hspace{-.4cm}&&p_0\gamma_+=(m_1+m_2)\gamma_0\gamma_+=m\gamma_0\gamma_+ \nonumber\\
\hspace{-.4cm}&&\hbox{and} \quad \gamma_+p_0=m\gamma_+\gamma_0,\label{4.47}
\end{eqnarray}
so (\ref{4.6}) simplifies to
\begin{equation}
\hbar \psi'\gamma_+=m\psi\gamma_0\gamma_+\gamma_1+q\lambda_e F\psi\gamma_+.\label{4.48}
\end{equation}
The problem now is to solve for $\psi'$. This is complicated by the fact that one cannot
divide out the null vector. The slickest way around this obstacle is as follows. First apply
(\ref{4.47}) to (\ref{4.48}) to get
\begin{equation}
\hbar \psi'\gamma_+p_0=-m^2\psi\gamma_-\gamma_1+q\lambda_e
F\psi\gamma_+ p_0.\label{4.49}
\end{equation}
Then use (\ref{4.46}) to get
$$\hbar \psi'p_0\gamma_+=\hbar[(\psi p_0)'-\psi p_0'\gamma_+]
=\hbar(p'\psi+p\psi'-\psi p_0')\gamma_+,$$
which, on inserting (\ref{4.48}) and the momentum equation (\ref{4.33}), becomes
\begin{multline}
\hbar \psi'p_0\gamma_+=\hbar\rho(q F\bdot u+\nabla\Phi)\psi\gamma_+\\
+p(p\psi\gamma_+\gamma_1+q\lambda_e F\psi \gamma_+)-\psi p_0'\gamma_+.\label{4.50}
\end{multline}
Adding (\ref{4.48}) and (\ref{4.50}), we obtain
\begin{multline}
2\hbar \psi'\gamma_+\bdot p_0
=\hbar\rho(qF\bdot u+\nabla\Phi)\psi\gamma_+\\
+\psi(p^2\gamma_+-m^2\gamma_-)\gamma_1
+q\lambda_e[2p\bdot  F\psi \gamma_+\\
+F\psi'(p_0\gamma_++\gamma_+p_0)]-\hbar\psi
p_0'\gamma_+,\nonumber
\end{multline}
This can be solved easily for $\psi'$ using
$$\gamma_+\bdot p_0=\frac{1}{2}(\gamma_+p_0+p_0\gamma_+)=p\bdot u=m,$$
and further simplified with
\begin{eqnarray}
\hspace{-.5cm}p^2\gamma_+-m^2\gamma_-&=&m[(m_1-m_2)\gamma_+-(m_1+m_2)\gamma_-]\nonumber\\
&=&mp_0\gamma_0\gamma_2,\nonumber\\
p_0'\gamma_+&=&(m_1'\gamma_0-m_2'\gamma_2)\gamma_+=m'\gamma_0\gamma_+.
\end{eqnarray}
Finally, we obtain the desired spinor equation:
\begin{multline}
\psi'=\frac{1}{2}\left[\frac{2}{\hbar}pe_0e_2e_1+\frac{q}{m_e}
F+\right.\\ \left.
\frac{1}{m}\left(\rho \nabla\Phi-\frac{q}{m_e}F\bdot
\pi-m'e_0\right)u\right]\psi,\label{4.52}
\end{multline}
where $\pi=p-m_ew=p-m_e\rho u$, as before.

Expressing (\ref{4.52}) in the general form,
\begin{equation}
\psi'\psi^{-1} =\rho'/\rho+\Omega=\dot{\rho} +\Omega,\label{4.52b}
\end{equation}
we can separately equate scalar and bivector parts.
The scalar part gives us
\begin{equation}
\hspace{-.8cm}\frac{\rho'}{\rho}+\frac{m'}{m}=\frac{(\rho m)'}{\rho m}=\frac{\rho}{m}
u\bdot  \nabla\Phi-\frac{q}{mm_e}F\bdot (p\wedge u),\label{4.53}
\end{equation}
which is identical to the equation (\ref{4.36}) that gave us the mass integral of motion
(\ref{4.37}). The bivector part gives us
\begin{multline}
\Omega=\frac{2}{\hbar}pe_0e_2e_1+\frac{q}{m_e}F\\
+\frac{1}{m}\left(\rho \nabla\Phi-\frac{q}{m_e}F\bdot
\pi\right)\wedge u+\frac{m'}{m}e_2e_0.\label{4.54}
\end{multline}
This is the general expression for rotational velocity that we were seeking.
It holds for any mass gauge choice, including $\rho=1$, which we now adopt for the balance of
this paper.

The first term in (\ref{4.54}) can be interpreted as a \textit{kinetic rotational velocity}:
\begin{equation}
\Omega_k\equiv \frac{2}{\hbar}pe_0e_2e_1=\frac{2}{\hbar}(m_1e_2-m_2e_0)e_1.\label{4.55}
\end{equation}
It is a spacelike vector (since $\Omega_k^2=-4p^2/\hbar^2$), so it generates a spacelike
internal \textit{zitter rotation}, as is evident from its contribution to the velocity
equation: $\Omega_k\bdot u=\omega e_1$, where the zitter frequency $\omega =2m/\hbar$ is
determined by the mass equation $m=m_e+\Phi$.
The second term $(q/m_e)F$ can be regarded as a generalization of the Lorentz force,
as it determines extrinsic bending of the particle history due to external fields (in
(\ref{4.32})), along with a torque on the spin (in (\ref{4.36})). The remaining terms
account for shift in the internal rotation rate to accommodate change in mass, but they do
not directly contribute to the equations for velocity and spin.

Further insight into the rotational velocity is obtained by multiplying (\ref{4.52}) by the
spin bivector (\ref{4.15b}). The $\dot{m}$ term cancels out (when $\dot{\rho}=0$), so only
the first two terms survive to give us the elegant relation
\begin{equation}
\hspace{-.4cm}\Omega S = \Omega \bdot S+\Omega\times S+\Omega\wedge S
=-p u+\frac{q}{m_e}F S. \label{4.62}
\end{equation}
 Separation into parts of homogenous grade yields:
\begin{equation}
\Omega \bdot S= -m +\frac{q}{m_e}F\bdot S=-m_e, \label{4.63}
\end{equation}
\begin{equation}
\Omega\times S=u\wedge p+\frac{q}{m_e}F\times S , \label{4.64}
\end{equation}
\begin{equation}
\Omega\wedge S=\frac{q}{m_e}F\wedge S. \label{4.65}
\end{equation}
Here we see that the mass integral of motion $m-\Phi=m_e=-\Omega \bdot S$ can be interpreted
as constraining projection of the spin onto the rotational velocity to be a constant of
motion.

It is noteworthy that (with $\rho=1$)
the mass integral of motion can be reformulated as an expression for the zitter frequency:
\begin{equation}
\omega=\omega_e +\frac{2}{\hbar}\Phi=\omega_e +\frac{q}{m_e} F\bdot(ue_1)
=\frac{d\varphi}{d\tau}, \label{4.66}
\end{equation}
where, anticipating further analysis, a \textit{zitter phase angle} $\varphi$ has been
introduced. We shall see that this angle is completely analogous to the phase angle in the
Dirac equation and, \textit{\'a fortiori}, to the phase angle in Schroedinger's equation.
Note that the phase angle may depend on the electromagnetic vector potential, but the
frequency shift is gauge invariant, as it has the form of a dynamical flux integral on the
circulating particle history. The present context also suggests that the phase $\varphi$ may
be a more fundamental time variable than the proper time.

Generalizing the free particle case (\ref{5.13b}), we identify
\begin{equation}
\lambda=\omega^{-1}=\frac{d\tau}{d\varphi}, \label{4.67}
\end{equation}
as the variable \textit{zitter radius}.
This interpretation is supported by the form of equation (\ref{4.32}), wherein
$\lambda$ can be identified as an intrinsic \textit{radius of curvature} for a helical particle
history.

The \textit{first curvature} $\kappa_1$ of a particle history measures the bending rate
orthogonal to its velocity. In the particle equation of motion (\ref{4.32}) the direction of
helical bending is given by the \textit{zitter vector} $e_1$. Hence, the first curvature is
given by
\begin{equation}
\kappa_1\equiv -\dot{u} \bdot e_1=\omega -\frac{q}{m_e}F\bdot (ue_1)=\omega_e=\lambda_e^{-1}.
\label{4.68}
\end{equation}
According to (\ref{4.39}) this quantity is rigorously constant!
However, it is composed of two parts: the first
 can be regarded as an \textit{intrinsic curvature} while the second is an ``extrinsic
curvature" due to external forces. This tells us that in response to
external forces the intrinsic curvature is adjusted to maintain an overall constant
value.
Its inverse is the curve's radius of curvature, which we recognize as the free particle
zitter radius.

\section{Analysis and Implications}

The formulation of our model for a lightlike electron with zitter is now complete.
In principle, the model can be applied to any problem in
quantum mechanics. The present discussion is limited to a general analysis of
solution techniques and a survey of promising possibilities for applications and
experimental test of the model.

\subsection{Zitter center}

An important unresolved issue is how best to define the relation between the timelike
reference history\, $x=x(\tau)$ and the lightlike particle history $z=z(\tau)$.
Relative to the instantaneous rest frame defined by $v=\dot{x}$, at any time
$\tau$ the two histories are separated by a displacement vector
\begin{equation}
r\equiv z-x. \label{4.70}
\end{equation}
If this is defined so that $|r|\approx \lambda_e$, we can regard the reference history as
specifying a \textit{center of curvature}, around which the helical particle revolves as time
progresses.
An essential condition on a \textit{zitter center history}  $x(\tau)$ is
\begin{equation}
\dot{x} \bdot u=\dot{x} \bdot \dot{z}=1,\label{4.71}
\end{equation}
as that defines the proper time.
What we need is a tractable equation of motion for $\dot{x}$ that satisfies these
constraints. This is akin to defining an equation of motion for the guiding center of a
helical orbit in plasma physics. Two possibilities present themselves immediately:
$\dot{x} =p/m_p$ or $\dot{x} =e_0$. Unfortunately, their equations of motion imply that they
both ``wobble'' with the zitter frequency, and if we try to smooth that out (below), there
is no guarantee that the history
$x(\tau)$ cannot drift outside the helix $z(\tau)$.

An alternative with attractive physical interpretation is to
 specify $x(\tau)$  by defining a zitter radius vector $r\equiv-\lambda e_1$ with
variable length specified by  (\ref{4.67}).
However, since the first curvature
$\kappa_1$ is a constant of the motion, a better choice may be $r_e\equiv-\lambda_e e_1$,
defining a zitter radius vector of fixed length.  The felicity of that choice is confirmed
by noting that it gives the spin potential (\ref{4.34}) the perspicuous form
\begin{equation}
\Phi=q F\bdot(u\wedge r_e).\label{4.72}
\end{equation}
As there is no crucial physical issue involved, we can defer choice among the
alternatives to be decided by convenience in specific applications.
The best choice may well depend on what approximations we adopt.

\subsection{Zitter averages and approximations}

As zitter fluctuations are so rapid, it is most convenient to separate them from zitter
means, which are more directly observable. The velocity $v=\dot{x}$ defines an
instantaneous rest frame for the electron at each time $\tau$, so we define the zitter mean
as an average over the free particle zitter period that keeps the \textit{zitter center
velocity} $v$ and the spin vector $s=(\hbar/2)e_3$ fixed.
With an overline to denote average value, basic zitter means are specified by
the expressions
\begin{equation}
\hspace{-.3cm}\obv=v=\dot{x} =\obu =e_0,\quad \obe_1=0=\obe_2,\quad \obe_3=e_3.
\label{6.1}
\end{equation}
Consequently, the mean of the \textit{spin bivector} is
\begin{equation}
 \obS = \frac{\hbar}{2}\overline{ue_1}=\frac{\hbar}{2}\overline{e_2 e_1}
=isv \,.\label{6.2}
\end{equation}
This approximation ignores variations in zitter radius and mass over a zitter period.

Since the electromagnetic field $F(z)=F(x+r)$ acts at the location of the particle, to get
its effect on the zitter center, we expand with respect to the zitter radius vector.
Accordingly, the average field at the zitter center is given by
\begin{equation}
\hspace{-.7cm}\obF(z)=F(x)+\overline{r}\bdot\nabla F(x)+\frac{1}{2}\overline{(r\bdot\nabla)^2}
F(x)+\ldots.
\label{6.3}
\end{equation}
Of course, $\overline{r}=0$, so to first order $\obF=F(x)$.

Neglecting second order effects gives us the mean mass
\begin{equation}
\obm=m_e+\overline{\Phi},\quad \hbox{with}\quad\overline{\Phi}
=\frac{q}{m_e} \obS\bdot F(x).\label{6.4}
\end{equation}
The mass shift $\overline{\Phi}$ should be recognized as a generalization of the Zeeman
interaction in atomic physics.

In the present approximation the mean momentum aligns with the zitter center velocity:
\begin{equation}
\obp=\obm_1\obe_0-\obm_2\obe_2=m_ev,
\label{6.4a}
\end{equation}
where we have used
\begin{equation}
\obm_1=m_e,\quad \hbox{so}\quad \obm_2=\obm-\obm_1=\overline{\Phi}.
\label{6.4b}
\end{equation}
Consequently,
the momentum equation (\ref{4.33}) can serve as an equation of motion for the zitter center.
Projecting out the effect of the mass derivative and averaging, we obtain from
(\ref{4.33}) a mean equation
\begin{equation}
m_e\dot{v}= q\obF\bdot v +v\bdot(v\wedge \nabla)\overline{\Phi}.
\label{6.6}
\end{equation}
This will be recognized  as the classical equation for a
charged particle with intrinsic spin, so  it can be regarded as the \textit{classical limit of
the zitter model}.
 With  $\obF=F(x)$, the first term on the right is the classical Lorentz force, while the
second is the Stern-Gerlach force.

For slowly varying electromagnetic fields there is not much difference between
equations (\ref{4.33}) and (\ref{6.6}), save that replacement of the Lorentz force
$qF\bdot v$ by $qF\bdot u$ produces a \textit{wobble} of $v$ about its mean value. The big
difference in the equations comes with electromagnetic fields oscillating with a
frequency close to the zitter frequency, for then resonance can occur, as we shall see.

Our average model is not complete without an equation of motion for the spin.
We can get that by computing the mean rotational velocity from (\ref{4.53}), with the
result
\begin{multline}
\overline{\Omega}=e_1e_2\omega_e
-\frac{1}{m_e}v\wedge\nabla\overline{\Phi}+\frac{q}{m_e}\obF\\
+e_0\left(e_1\frac{2\overline{\Phi}}{\hbar}-e_2\frac{\dot{\overline{\Phi}}}
{\overline{m}} \right) .
\label{6.7}
\end{multline}
The last pair of terms might be regarded as a second order effect, but if the zitter vectors
$e_1, e_2$ are "dotted out" it surely becomes first order.  The mean motion of the comoving
frame is now given by
\begin{equation}
\dot{e}_\mu =\overline{\Omega}\bdot e_\mu,
\label{6.8}
\end{equation}
which determines the mean spin motion as well.
Let's call this the \textit{minimal zitter model,} because it is
the simplest model that captures the main features of zitterbewegung.
Obviously, the model can be extended by including higher order moments in the
averaging.

In the next Section we see that the minimal model corresponds closely
to Dirac theory, with $v =\obu$ corresponding to the direction of the Dirac current, while
$\obS$ matches the Dirac spin.
That may be surprising, since the spin bivector here
satisfies the null constraint $S^2=0$, and no such thing appears in standard
accounts of Dirac theory. The discrepancy is resolved by considering the \textit{spacetime
split} \cite{Hest03b}:
\begin{eqnarray}
 S&=&\frac{\hbar}{2}ue_1=\frac{\hbar}{2}isu=Sv^2 \nonumber\\
 &=&(S\bdot v+S\wedge
v)v=(m_er+is)v\,,\label{6.9}
\end{eqnarray}
where the zitter radius vector and mean spin are given by
\begin{equation}
m_er\equiv-\frac{\hbar}{2}e_1=S\bdot v
\end{equation}
and
\begin{equation}
\obS = (S\wedge v)v=isv
\,,\label{6.10}
\end{equation}
with the sign chosen so the radius vector points away from the zitter center.

This is a split of the spin $S$ into a rapidly rotating part $m_erv$ and a slowly
precessing part $\obS$.
 The specification of spin vector here is algebraically identical to
the one in Dirac theory \cite{Hest03b}.
The rapidly rotating unit vector $\hat{r}=-e_1$ is
also inherent in Dirac theory, but its physical significance has been overlooked.
In Dirac theory, identification of the radius vector and the spin vector as parts of the
null bivector $S$ has not been made heretofore. The implication
is that the spin observable in Dirac theory is only the zitter average $\obS$ of the full
spin $S$. Nevertheless, zitter is still present in the Dirac equation as a rotating
phase factor in the wave function, and we shall see that there are indeed possibilities
to detect it.

The ``$v$-split" of the spin in (\ref{6.9}) entails a $v$-split of the
electromagnetic dipole moment into electric and magnetic parts:
\begin{eqnarray}
\frac{q}{m_e} S &=& \d_v+i\bmu_v \quad \hbox{with}\nonumber\\
\d_v &=& -qrv, \quad \bmu_v= \frac{q}{m_e}sv=\frac{q}{m_e}\obS\,.\label{6.11}
\end{eqnarray}
Note that the electric dipole  has constant magnitude
$\pm |\d_v|=q\lambda_e=q\hbar/2m_e$, so it only rotates.
A corresponding $v$-split of the external electromagnetic field has the form
$F=\E_v+i\B_v$, where electric and magnetic parts are given by
\begin{equation}
\E_v=\frac{1}{2}(F-vFv), \qquad i\B_v=\frac{1}{2}(F+vFv). \label{6.12}
\end{equation}
Of course, these are electric and magnetic fields as ``seen" in the
instantaneous rest frame of the zitter center, not to be confused with fields in a ``lab
frame."

The induced mass shift can now be expressed in the
physically perspicuous form
\begin{equation}
\Phi=\frac{q}{m_e}S\bdot F= \d_v \bdot\E_v-\bmu_v \bdot \B_v . \label{6.13}
\end{equation}
This is a general result, holding exactly in every application of the zitter model.
As $\d_v$ is a rotating vector, this result is consistent with the well established
experimental fact that the electron does not have a static electric dipole moment.
Of course, if the $\E_v$ field is slowly varying, the effective shift is reduced to
the Zeeman interaction  $\overline{\Phi}=-\bmu_v \bdot \B_v$.

However, the spin split described here has a drawback, namely, that the reference
direction $v$ is constantly changing with time, so it is difficult to compare spin
directions at different times.
We show next how that drawback can be eliminated by examining the spin split
more thoroughly from a different point of view.

\subsection{Zitter and spin in the electron rest frame}
Transformation of the time dependent instantaneous rest frame specified by velocity
$v=v(\tau)$  to an inertial rest frame specified by a constant vector
$\gamma_0$ is a \textit{boost}, specified by
\begin{eqnarray}
v&=&L\gamma_0\tL=L^2\gamma_0 \qquad \hbox{or}\nonumber\\
L^2&=&v\gamma_0=v_0(1+\v), \label{6.14}
\end{eqnarray}
where $v_0=v\bdot\gamma_0$ is the relativistic time dilation factor relating particle
proper time $\tau$ to inertial time $t$, that is,
\begin{equation}
v_0\equiv v\bdot \gamma_0=\frac{1}{(1+\v^2)^\half}=\frac{dt}{d\tau},\qquad
\v=\frac{d\x}{dt}\label{6.15}
\end{equation}
is the \textit{relative velocity} of the particle in the inertial reference frame.
Let us refer to this frame as \textit{the electron rest frame,} though we shall generalize the
definition somewhat in a later subsection

We can solve (\ref{6.14}) to get an explicit expression
for the rotor $L$ that generates the boost \cite{Hest03b}:
\begin{equation}
L=(v\gamma_0)^{\half}=\frac{1+v\gamma_0}{[2(1+v_0)]^\half}
=\frac{1+v_0+v_0\v}{[2(1+v_0)]^\half}. \label{6.16}
\end{equation}
However, it is usually easier to work with the simpler form for $L^2$ in (\ref{6.14}).

The deboost of the $v$-split for spin $S$ in (\ref{6.9}) to a split in the electron rest
frame is now given by
\begin{equation}
 S=LS_0\tL  \qquad \hbox{where}\qquad  S_0=-m_e\r+i\s,\label{6.17}
\end{equation}
with electric and magnetic moments defined by
\begin{equation}
\frac{q}{m_e}S_0=-q\r+i\frac{q}{m_e}\s\equiv \d+i\bmu.\label{6.18}
\end{equation}
The deboost of the particle velocity has the form
\begin{equation}
 uv=L(1+\u)\tL.  \label{6.18a}
\end{equation}
Since spin and velocity are related by $Su=0$, from (\ref{6.17}) and (\ref{6.18a}) we
obtain
\begin{equation}
 S_0(1+\u)=(-m_e\r+i\s)(1+\u)=0.\label{6.18b}
\end{equation}
Separating parts of homogeneous grade, we ascertain
\begin{equation}
\r\bdot \u=0, \qquad \s\bdot \u=0 \label{6.18d}
\end{equation}
and
\begin{equation}
\s=-im_e\r\u=m_e\r\btimes \u.\label{6.18c}
\end{equation}
This makes spin look like a classical orbital angular momentum. However, the velocity
$\u=\dot{\z}$ is not generally the derivative of the radius vector $\r$, as we shall see.

It is often useful to express the relation of spin to zitter in terms of a
rotating orthonormal frame:
\begin{equation}
 \e_k\equiv \tL e_kv L=U\bsig_k\tU,\label{6.18f}
\end{equation}
where $\e_1=\hat{\r}$, $\e_2=\u$ and $\e_3=\hat{\s}$.
However, it is physically more enlightening to first study the dynamics of spin and zitter
more directly.

To treat interaction with external fields in the electron rest frame, we need
the deboost of an electromagnetic field:
\begin{equation}
F_0\equiv \tL FL=\E_0+i\B_0,\label{6.18e}
\end{equation}
where
\begin{equation}
F=\E+i\B= LF_0\tL\label{6.19}
\end{equation}
expresses the field in terms of electric and magnetic fields defined in the lab frame.

Note that
\begin{eqnarray}
&&\hspace{-1.1cm}<SF>=<LS_0\tL F>=<S_0\tL FL>\nonumber\\
&=&<S_0F_0>=<(\d+i\bmu)(\E_0+i\B_0)>.
\end{eqnarray}
Whence the spin potential (= mass shift) assumes the form
\begin{equation}
\hspace{-.2cm}\Phi=\frac{q}{m_e}S\bdot F= \frac{q}{m_e}S_0\bdot F_0=
\d \bdot\E_0-\bmu \bdot \B_0 .\label{6.20}
\end{equation}
However, this is not the final form required for physical interpretation.

We still need explicit expressions for the deboosted fields $\E_0$ and
$\B_0$ in terms of lab fields $\E$ and $\B$.
Using (\ref{6.14}) we have, with obvious notation,
\begin{equation}
F_0=\tL FL=F_\parallel+F_\perp L^2\label{6.21}
\end{equation}
where
\begin{equation}
F_\parallel \v=\v F_\parallel,\quad F_\perp\v=-\v F_\perp.\label{6.21a}
\end{equation}
Whence
\begin{equation}
\E_0=\E_\parallel+v_0\E_\perp +v_0\v \btimes \B,\label{6.22}
\end{equation}
where
\begin{equation}
\E_\parallel=(\E \bdot \hat{\v}) \hat{\v},
\quad \E_\perp =\E-\E_\parallel.\label{6.22a}
\end{equation}
Similarly,
\begin{equation}
\B_0=\B_\parallel+v_0\B_\perp -v_0\v \btimes \E.\label{6.23}
\end{equation}
Insertion into (\ref{6.20}) gives explicit expressions for the interaction of the
electron's electric and magnetic dipoles with external fields.

Physical interpretation of spin dynamics is facilitated by transforming the spin equation
of motion (\ref{4.36}) to the instantaneous rest frame. Thus,
\begin{equation}
\hspace{-.6cm}\dot{S}=L(\dot{S}_0+\Omega_0\times S_0)=L(m\e_2+\frac{q}{m_e}F_0\times S_0)\tL
,\label{6.24}
\end{equation}
where
the rotational velocity $\Omega_0$ is determined by differentiating the boost
 (\ref{6.16}), with the result \cite{Hest03b}:
\begin{eqnarray}
 L\Omega_0\tL&=&\Omega_v \equiv 2\dot{L}\tL=\frac{\dot{v}\wedge(v+\gamma_0)}{1+v_0}
 \nonumber\\
&=&v_0\dot{\v}+\frac{\dot{v_0}\v}{1+v_0}
+i\frac{v_0^2\dot{\v}\btimes \v}{1+v_0}.
\label{6.25}
\end{eqnarray}
As is evident in equation (\ref{6.24}), this quantity acts like an
effective electromagnetic field induced by acceleration of the rest frame. The
magneticlike term at the right side of the equation is responsible for the classical
\textit{Thomas precession} of the spin.

Inserting the split (\ref{6.17}) with (\ref{6.18}) into (\ref{6.24}), we have
\begin{multline}
\dot{S}_0=m_e\dot{\r}+i\dot{\s}\\=m\e_2+(F_0-\frac{m_e}{q}\Omega_0)\times (\d+i\bmu)
.\label{6.26}
\end{multline}
Now we introduce the split
\begin{equation}
 \a+i\b \equiv F_0-\frac{m_e}{q}\Omega_0=\tL(F-\frac{m_e}{q}\Omega_v)L,\label{6.27}
\end{equation}
noting that expressions for $\a$ and $\b$ can be obtained from (\ref{6.25}) by the deboost
(\ref{6.21}).
Using
\begin{multline}
(\a+i\b)\times (\d+i\bmu)\\=\a\wedge\d-\b\wedge\bmu+i(\a\wedge\bmu+\b\wedge\d)
\label{6.29}
\end{multline}
with $\a\wedge\d=i(\a\btimes \d)$, we split (\ref{6.26}) into coupled equations of
motion for zitter and spin:
\begin{equation}
m_e\dot{\r}=m\e_2+\bmu\btimes\a+\d\btimes\b,\label{6.30}
\end{equation}
\begin{equation}
\dot{\s}=\a\btimes\d+\bmu\btimes\b.\label{6.31}
\end{equation}
These equations are helpful for analyzing the dynamical behavior of spin and zitter
vectors.

The spin equation (\ref{6.31}) is most familiar. Its last term $\bmu\btimes\b$ is
recognized as the usual spin precession torque, including the Thomas precession, as
already noted. The term $\a\btimes\d$ is something new.
First note that
\begin{equation}
\dot{\s}\bdot \s=(\a\btimes\d)\bdot \s=(\d\btimes\s)\bdot \a =0=\e_2\bdot \a .\label{6.32}
\end{equation}
To prove that this quantity does indeed vanish, we use (\ref{6.27}):
\begin{multline}m\a\bdot\e_2=<L\a\tL u\wedge p>=<(F-\frac{m_e}{q}\Omega_v)u\wedge p>\\
=\frac{m_em}{q}[\dot{u}\bdot v+\dot{v}\bdot u]=0.
\end{multline}
It follows from this constraint that we can write
\begin{equation}
\a\btimes\d=\hat{\s}\btimes\d\,(\hat{\s}\bdot \a).\label{6.33}
\end{equation}
This torque has several interesting properties. First, it rotates $\s$ about the vector
$\e_2$, which is the direction of zitter circulation in the rest frame. Second, the
torque decreases in magnitude until it vanishes at $\s\bdot \a=0$. Third, its zitter
average is zero if $\a$ is slowly varying.
These properties suggest that the term is a good candidate for a spin polarization
torque.

The zitter equation (\ref{6.30}) has similar properties, of course, with the additional
term $m\e_2$ expressing the high frequency zitter rotation.
We only note that the $\bmu\btimes\a$ torque ensures that $\r$ rotates around $\e_2$ along
with $\s$.

\subsection{Zitter model for a static potential}

The interaction of atoms and crystals with an electron is usually modeled with a static
potential.
For that purpose we introduce a static vector potential defined in the inertial frame of
$\gamma_0$ at each time $t$ and position $\x$ by
\begin{equation}
 qA=V\gamma_0 \qquad \hbox{where} \qquad V= V(\x). \label{6.34}
\end{equation}
This determines an electromagnetic field
\begin{equation}
\hspace{-.6cm}qF=\nabla V \wedge \gamma_0 =-\gamma_0 \nabla V=-\gamma_0\wedge \nabla V\equiv-\bnabla V.
\label{6.35}
\end{equation}
This, in turn, determines an electromagnetic force on the particle
\begin{equation}
\hspace{-.6cm}qF\bdot u=\nabla V u\bdot \gamma_0 -\gamma_0 \dot{V},
\quad \hbox{where} \quad \dot{V}=u\bdot \nabla V.\label{6.36}
\end{equation}
Inserting this into the momentum equation (\ref{4.33}) we obtain
\begin{equation}
\frac{d}{d\tau}(p+V\gamma_0)=-\nabla(Vu\bdot\gamma_0+\Phi).\label{6.37}
\end{equation}
Next, we introduce the space-time split \cite{Hest03b}
$p\gamma_0=p\bdot\gamma_0+p\wedge\gamma_0\equiv p_0+\p$ and separate spatial and
temporal parts. This yields a \textit{ conserved energy}:
\begin{equation}
E=p_0+V\label{6.38}
\end{equation}
and, after factoring out $v_0\approx u\bdot\gamma_0$, a momentum force law
\begin{equation}
\frac{d\p}{dt}=-\bnabla(V +v_0^{-1}\Phi),\label{6.39}
\end{equation}
where $v\gamma_0=v_0(1+\v)$ with $\v=d\x/dt$ and $\p=mv_0\v$.

We still need an explicit form for the spin potential. From (\ref{6.36}) we get
\begin{equation}
\hspace{-.3cm}\Phi= \lambda_e qF\bdot (ue_1) = u\bdot \gamma_0\,e_1\bdot \nabla V
-\gamma_0 \bdot e_1\,\dot{V}. \label{6.40}
\end{equation}
However, it is better to use (\ref{6.22}) and (\ref{6.23}) in (\ref{6.20}) to get the form
\begin{eqnarray}
&&\Phi= q\lambda_e[\e\bdot (\E_\parallel+v_0\E_\perp)+\hat{\s}\bdot(v_0\v \btimes \E)]
\nonumber\\
&&\hbox{with} \qquad q\E=-\bnabla V, \label{6.41}
\end{eqnarray}
where  $\e$ denotes the unit zitter vector.
This has the advantage of clearly separating the high frequency zitter from the low
frequency spin contributions, and it exhibits the zitter radius $\lambda_e$ as governing
the strength of the interaction.

These results can be applied to experimental search for observable effects of electron
zitter. As the energy conservation law (\ref{6.38}) is identical to the usual one,
zitter will be manifested only in momentum fluctuations, which are so small and rapid that
they are observable only in resonances.
Consider \textit{Mott scattering} by a Coulomb field, for example.
In low and high energy ranges the cross section will not be significantly affected by
the very high frequency zitter rotation, so the standard result should be
obtained \cite{Bjorken64}. However, in an intermediate range where the impact parameter
is on the order of a zitter wavelength, the zitter structure of the particle trajectory
should have a significant effect on the scattering. How big an effect awaits
calculation!

As explained below, most resonances involve zitter
field theory, which is beyond the scope of the present treatment. However, a new possibility
for amplifying zitter resonances in crystals has recently appeared in  electron channeling
experiments, to which we now turn.

\subsection{Zitter resonance in electron channeling}

When a beam of electrons is channeled along a crystal axis, each electron is subject to
periodic impulses from atoms along the axis. When the energy of the beam is adjusted so the
crystal period  matches the period of the electron's zitter dipole, a resonant
interaction may be expected to alter the distribution of transmitted electrons. Indeed, a
pioneering experiment in search of such a resonance has already been performed, but without
knowledge of the dipole interaction mechanism described here. The anticipated
resonance was observed at close to the de Broglie frequency, which is precisely half
the zitter frequency \cite{Gouanere05}.  Our purpose here is to show how this result
 can be explained quantitatively by the zitter model.
This confluence of theory and experiment
provides ample grounds for repeating the experiment with greater accuracy to confirm the
results and look for details suggested by the theory.

\subsubsection{Experimental specifications}

 The anticipated resonant energy is easily calculated from the de Broglie's (circular) frequency
$\omega_B=m_ec^2/\hbar$. One of de Broglie's original insights was that the frequency of a moving
electron observed in a laboratory will be $\omega_L=\omega_B/\gamma$, where $\gamma=v_0$ is the
relativistic time dilation factor.
The distance traversed during a clock period is $d=2\pi c\beta/\omega_L=hp/(m_ec)^2$.
For the silicon crystal used in the experiment, the interatomic distance along the $<110>$
direction is $d=3.84\,\AA$, which implies a resonant momentum $p=80.874$ MeV/c.

In channeling the maximum crystal potential is a few hundred electron volts at most, so
in the 80 MeV region of interest to us, the effective electron mass $M\equiv\gamma m_e =E/c^2$ is
constant to an accuracy of $10^{-5}$, and $\gamma=158$.

In axial channeling electrons are trapped in orbits spiraling around a crystal axis. To a first
approximation, the crystal potential can be modeled as the potential for a chain of atoms, so it
has the form
\begin{equation}
 V(\x)=V(r,z)=U(r)P(2\pi z/d) ,\label{21}
\end{equation}
where $\x(t)=\r+z\bsig_z$ is the particle position from the first atom in the chain, with
$r=|\r|$. The longitudinal potential $P(2\pi z/d)= P(\omega_0 t)$ is periodic with a \textit{tunable
frequency} $\omega_0=2\pi\dot{z}/d$ that varies with the energy $E$. Note that at the
expected resonance $\dot{z}=d/T_L$, so $\omega_0=2\pi/T_L=\omega_B/\gamma$
corresponds to the de Broglie frequency.

Our problem is to calculate perturbations on the transverse component of the momentum
vector, as that can remove electrons from stable orbits in the beam. The transverse
component of equation (\ref{6.39}) has the familiar form of a nonrelativistic equation:
\begin{equation}
 M\ddot{\r}=-\hat{\r}(P\partial_r U+
\gamma^{-1}\partial_r \Phi),\label{22}
\end{equation}
while the longitudinal component has the form
\begin{equation}
 M\ddot{z}=-(U\partial_z P+
\gamma^{-1}\partial_z \Phi),\label{22a}
\end{equation}
where now the overdot indicates differentiation with respect to ``lab time'' $t$.
In the energy range of interest, it is easy to show that oscillations
in the transverse velocity can be ignored, so we regard $\dot{z}$ as a constant tunable
velocity close to the speed of light as already assumed above.

\subsubsection{Crystal potential}

To proceed further, we need an explicit model of the crystal potential.
For analytic simplicity, we approximate the
potential  by the first two terms in a Fourier expansion with respect to the
reciprocal lattice vector. So write (\ref{21}) in the more specific form
\begin{equation}
V=U(r)(1+\cos\omega_0t),\label{22b}
\end{equation}
where the coefficient of the second
term is set to make the potential vanish between atoms.  The first term is the potential for a
uniformly charged string, which  (in its simplest version) has Lindhard's
form \cite{Gemmell74,Lindhard65}:
\begin{equation}
 U(r)=-k\ln [1+(Ca/r)^2],\label{23}
\end{equation}
where $Ca=0.329 \AA$, and (with $Z=14$ for silicon) the coupling constant $k= Ze^2/d =52.5$ eV
is the product of electron charge with charge per unit length of the string.
The constant $a=0.885Z^{-1/3}a_0=(0.885)(14)^{-1/3}(0.529\AA)=.190\AA$ is the Fermi-Thomas
screening radius, and the constant $C^2=3$ is a fairly accurate fit over the range of
interest.

The string potential is defined as an average over atomic potentials; thus,
\begin{equation}
 U(r)=\int^\infty_{-\infty} V_{\text{atom}} [(r^2+z^2)^{1/2}]\frac{dz}{d},\label{23a}
\end{equation}
where the screened atomic potential is given by
 \begin{equation}
 V_{\text{atom}}(R)=-\frac{Ze^2}{R}\varphi(R/a).\label{23b}
\end{equation}
To get (\ref{23}), Lindhard used the screening function
 \begin{equation}
 \varphi(R/a)=1-[1+(Ca/R)^2]^{-1/2}.\label{23c}
\end{equation}
Actually, the experiment is not very sensitive to the shape of the potential, so long as it is
sufficient to bind the electron to an orbit around the crystal axis.
In the first approximation $V=U(r)$ the projection of the electron orbit onto a transverse plane
 looks like a precessing ellipse or rosette.  The second term $U\cos\omega_0t$ is ignored in most
channeling calculations, as it merely produces small harmonic oscillations on the radius of the
precessing ellipse.   However, the periodicity of the second term is essential for resonance in the
zitter perturbation of interest here.

\subsubsection{Classical channeling orbits}

Ignoring the zitter perturbation in (\ref{22}) for the time being, we
seek to ascertain the effect of the periodic factor $P(\omega_0t)=1+\cos\omega_0t$ on the orbital
motion.
With an obvious change of notation, we can represent the radius vector in the complex form
$\r=re^{\i\theta}$, where the imaginary $\i$ is the bivector generator of rotations in the
transverse plane. Then equation (\ref{22}) assumes the complex form
\begin{equation}
 M\ddot{\r}=-U'Pe^{\i\theta}=-\frac{U'}{r}(1+\cos\omega_0t)\r,\label{24}
\end{equation}
with $U'\equiv \partial_rU$.

We are interested only in radial oscillations, so we use conservation of angular
momentum $L=Mr^2\dot{\theta}$ to separate out the rotational motion.
With the periodic driving factor omitted, equation (\ref{24}) admits the energy integral
\begin{eqnarray}
 &&E_\perp =\frac{1}{2}M\dot{r}^2 + W(r)\nonumber\\
 &&\hbox{where}\qquad
W(r) =\frac{L^2}{2Mr^2} +U(r).\label{24a}
\end{eqnarray}
Let us expand this around a circular orbit of radius $r_0$, and for quantitative estimates take
$r_0=0.50 \AA$ as a representative intermediate radius. For $\r_0=re^{\i\theta_0}$, equation
(\ref{24}) gives us
\begin{equation}
\dot{\theta_0}^2=\frac{U_0'}{Mr_0} =\frac{L^2}{M^2r_0^4},\label{25}
\end{equation}
where $U_0'=U'(r_0)$. In terms of $x=r-r_0$, expansion of (\ref{24a}) gives us
\begin{equation}
 E_\perp =\frac{1}{2}M\dot{x}^2 + W_0 +\frac{1}{2}W_0''x^2,\label{26}
\end{equation}
where
\begin{eqnarray}
  &&\hspace{-.5cm}W_0= U_0 +r_0U_0'/2,\quad W_0'= U'_0 -U_0'=0,\nonumber\\
  &&\hspace{-.5cm}W_0''= U''_0 +3U_0'/r_0
,\label{27}
\end{eqnarray}
with\quad $U_0=-18.9$ eV,
\begin{equation}
 U' = \partial_rU=\frac{2k}{r}\left[\frac{(Ca/r)^2}{1+(Ca/r)^2}\right],\nonumber
\end{equation}
with\quad $r_0U_0'=31.7$ eV,\quad and
\begin{equation}
U'' =-\frac{U'}{r}\left[\frac{3+(Ca/r)^2}{1+(Ca/r)^2}\right],\nonumber
\end{equation}
with\quad $r_0^2U_0''=-76.0$ eV.

Differentiating  (\ref{26}) and reinserting the periodic driving
factor, we obtain the desired equation for radial oscillations:
\begin{equation}
\ddot{x} +\Omega_0^2(1+\cos\omega_0t)x=0 ,\label{28}
\end{equation}
where, for mass $M$ at the expected resonance,
\begin{multline}
\Omega_0^2\equiv \frac{W_0''}{M}=\frac{3U_0'+r_0U_0''}{Mr_0}\\
=\frac{3\times 31.7-76.0}
{80.9\times10^6}\left(\frac{c^2}{r_0^2}\right),\label{29}
\end{multline}
so $\Omega_0=4.21\times10^{15}\,\rm{s}^{-1}$.
This should be compared with the
$\dot{\theta_0}=4.75\times10^{15}\,\rm{s}^{-1}$ from (\ref{25})
and the expected resonant frequency $\omega_0=\omega_B/\gamma=4.91\times 10^{18}\,\rm{s}^{-1}$.
We note that the distance traveled in one orbital revolution is
$d_r=2\pi c/\dot{\theta_0}=3.97\times 10^3 \AA=0.397\rm{\mu m}$. Thus, the orbit makes 2.52
revolutions in passing through the one micron crystal, so the orbital
revolutions are only weakly coupled to the high frequency radial oscillations.

Equation (\ref{28}) is a special case of \textit{Mathieu's equation}:
\begin{equation}
\ddot{x} +q(1+h\cos\omega t)x=0 .\label{30}
\end{equation}
According to \textit{Floquet's Theorem}, this equation has solutions of the general form
\cite{Morse53,Stegun72}
\begin{equation}
x(t)=e^{st}\sum_{n=-\infty}^{\infty}a_ne^{i\,n\,\omega t}.\label{31}
\end{equation}
Substitution into (\ref{30}) yields a three term recursion relation that can be solved for
$s$ and ratios of the Fourier coefficients $a_n$. In general, the \textit{Floquet exponent}
$s=s(q,h)$ is a complex constant, so it determines whether solutions are unstable or periodic.

Since $\Omega_0 << \omega_0$ in equation (\ref{28}), its Floquet exponent $s=i\Omega$ is pure
imaginary, and its recursion relations give
\begin{eqnarray}
&&\hspace{-.5cm}\Omega^2=\frac{3}{2}\Omega_0^2, \qquad \hbox{and}\nonumber\\
&&\hspace{-.5cm}\frac{a_n}{a_{n-1}}=\frac{a_{-n}}{a_{-n+1}}=\frac{\Omega_0^2}{2n^2\omega_0^2}
\quad \hbox{for}\,n\geq 1.\label{32}
\end{eqnarray}
Therefore, the solution is dominated by the first order term, with the particular
form
\begin{eqnarray}
&&\hspace{-.5cm}x(t)=a(\cos\Omega t)\cos\omega_0t=\frac{a}{2}(\cos\omega_+ t+\cos\omega_- t),
\nonumber\\
&&\hspace{-.5cm}\omega_\pm =\omega_0\pm \Omega.\label{33}
\end{eqnarray}
As might have been anticipated, this describes a harmonic oscillator with high frequency
$\omega_0$ and a slowly varying amplitude with frequency $\Omega$, which is equivalent to a sum
of two oscillators with frequencies $\omega_{\pm}$ separated by $\omega_+ -\omega_-=2\Omega$.

We shall see that, at the resonant frequency $\omega_0=\omega_B/\gamma$, the frequency shift
$\Omega=(0.857\times 10^{-3})\omega_0$ is the right order of magnitude to contribute to
experimental effects. Moreover, this quantity has been estimated at the particular radius
$r=0.50 \AA$, and it may be larger by an order of magnitude for smaller radii of experimental
relevance. Accordingly, a distribution of $\Omega$ values will contribute to the experiment.

A limit on the maximum radius of a channeled electron is set by the requirement that $W(r)$
in (\ref{24a}) must be negative for a bound orbit. A sharper limit is set by neighboring atoms.
The total crystal potential can be modeled as a sum of chain potentials isomorphic
to the one for the channeling axis. For larger radii perturbation from other chains can induce
transition to a neighboring chain, with the result that channeled electrons jump randomly from
chain to chain. We are interested in this effect only to the extent that it affects
the distribution of electrons transmitted by a single channel. However, a quantitative estimate
 of such transitions will not be attempted here.

In silicon, the closest chain to a $<110>$ channel is at a distance of $1.36 \AA$ with atoms
located at positions $z_{n+1/2}=(n+1/2)d$ alternating with the positions $z_n=nd$ along the
channel. Considering the slow precession of a channeled orbit at the resonant frequency, this
chain will resonate with it for hundreds of atomic steps. In fact, the interaction might lock
onto the orbit to prevent precession during resonance. This may indeed have a substantial
effect on channeled electrons, but we will not investigate it further here.

\subsubsection{Zitter perturbations}

Now we are prepared to consider the effect of zitter perturbations on the orbit.
To evaluate the zitter potential $\Phi$, we use equation (\ref{6.41})
with $\E_\parallel=(\E\bdot \hat{\v})\hat{\v}$ and $\hat{\v}\approx \bsig_z=\bnabla z$,
and we ignore the spin term (because it vanishes on averaging over spin
directions). By the way, we don't need to use the spin equation of motion
(\ref{4.36}) in our calculations; we only need the fact that it implies that the unit zitter vector
$\e$ rotates rapidly in a plane that precesses slowly with the spin $\s$.

Inserting the electric potential into (\ref{6.41}), we get
\begin{equation}
\Phi=-\lambda_e[\gamma\e\bdot \hat{\r} U'P +\e\bdot \bsig_zUP'].\label{54}
\end{equation}
Since $\hat{\r}$ is effectively constant over a zitter period, we can make the zitter
oscillations explicit by writing $\e\bdot
\hat{\r}=\cos(\omega_Z t/\gamma+\delta)$ and $\e\bdot \bsig_z= \sin(\omega_Z
t/\gamma+\delta')$, where $\omega_Z$ is the zitter frequency. Thus,
\begin{multline}
\Phi=-\lambda_e[\gamma U'P\cos(\omega_Z t/\gamma+\delta) \\
+UP'\sin(\omega_Z t/\gamma+\delta')]
,\label{55}
\end{multline}
where the smaller second term can be neglected.

The spin potential has a twofold effect on electron motion: first, as a  shift in zitter frequency
according to equation (\ref{4.66}); second, as a perturbation of the momentum in equation
(\ref{22}).  For the frequency shift we get the explicit expression
\begin{equation}
\omega_Z=\omega_e-\frac{\gamma U'}{m_ec}\cos(\omega_et/\gamma+\delta).\label{56}
\end{equation}
The modulus of the oscillating term has the estimated mean value
$\gamma U'c/m_ec^2=(158)(31.7\rm{eV}/0.5\AA)(3\times 10^{18}\AA\rm{s}^{-1}/(0.511\times
10^6\rm{eV})=1.96\times 10^{16}\rm{s}^{-1}$. Compared to
$\omega_e=1.55\times10^{21}\rm{s}^{-1}$, this quantity is too small by
$10^{-3}$ to play a role in the present experiment. However, the value of $U'$, estimated here at
a radius of $r_0=0.5\AA$, may be regarded as lower bound, as the logarithmic potential is a very
soft approximation to a realistic potential, which is much sharper at the screening radius
$a=0.190\AA$ where the logarithmic potential is invalid.  We shall see that close resonant
encounters with atomic nuclei play a dominant role in the experiment, so the
oscillating term might indeed contribute a mean frequency shift on the order of $10^{-2}\omega_e$,
which could show up in experimental data. Replacement of the soft periodic modulus
$P(\omega_0t)= (1+\cos\omega_0t)$ by one that is more sharply peaked at the atomic
sites may strengthen this conclusion. That being said, for the rest of our analysis, it suffices to
assume $\omega_Z=\omega_e$.

Inserting (\ref{55}) into (\ref{22}), we get an explicit expression for the zitter perturbation
term:
\begin{eqnarray}
-\gamma^{-1}(\bnabla\Phi)_\perp&=&-\gamma^{-1}\hat{\r}\,\partial_r \Phi
\nonumber\\
&=&\lambda_e\hat{\r}\,U''P\cos(\omega_Z t/\gamma+\delta) .
\label{57}
\end{eqnarray}

\subsubsection{Zitter resonance}

The task remains to show that this interaction can produce a resonant amplification of the
electron's orbit. Inserting it into equation (\ref{22}) with a convenient choice of phase and
writing
$\omega=\omega_Z/\gamma$, we get the
 equation of motion
\begin{equation}
 M\frac{d^2\r}{dt^2}=-\r \frac{U'}{r}(1+\cos\omega_0 t)(1+\frac{\lambda_e}{R}
\cos\omega t),\label{58}
\end{equation}
where $R\equiv -U'/U''=r[1 +(Ca/r)^2]/[3 +(Ca/r)^2]$ is an \textit{effective (screened) radius}.
Ignoring the amplitude modulation as determined in (\ref{33}), we can reduce this to a radial
equation
\begin{equation}
 \ddot{x}+\omega_0^2(1+\frac{\lambda_e}{R}
\cos\omega t)x=0.\label{59}
\end{equation}
Of course, we can replace $\omega_0$ in this equation by $\omega_{\pm}$ to get two separate
resonant peaks.

For $h=\lambda_e/R$ constant, (\ref{59}) is Mathieu's equation (\ref{30}), so let us solve it
using
$h=\lambda_e/R_0=1.931\times10^{-3}/0.208=9.283\times 10^{-3}$ as an approximation. For small
values of $h$ such as this, equation (\ref{59}) has a first order resonance at $\omega=2\omega_0$.
An easy way to see that is to regard (\ref{59}) as an equation for a driven harmonic oscillator
with driving force proportional to its amplitude.
In the experiment, the ``natural frequency'' $\omega_0$ was adjusted by
continuously varying the initial energy of the electron until a resonance was found.
	As for a periodically driven harmonic oscillator, resonance occurs when the driving
frequency equals the natural frequency. In this case, that means
$\omega-\omega_0 =\omega_0$ or $\omega_0 =\omega/2=\omega_Z/2\gamma$,
which explains why the resonance occurs at half the zitter frequency --- a surprising
result until contemplation shows that it is an obvious feature of parametric resonance!

More generally, it can be shown that (\ref{59}) has resonances at
$2\omega_0=n\omega$, for
$n=1,2,.\,.\,.$.  We demonstrate that explicitly for the first order resonance by truncating the
series in  (\ref{31}) to the form \cite{Landau69}
\begin{equation}
x=e^{st}[a\cos\frac{\omega t}{2}+b\sin\frac{\omega t}{2}].\label{60}
\end{equation}
For a resonant solution, the Floquet exponent $s$ must be real and positive. [Actually, to get
this form from (\ref{31}) we should use the Floquet coefficient $s-i\omega/2$ and incorporate
its imaginary part in the series. Thus, there is some ambiguity in the definition of Floquet
exponent.]

To validate the trial solution and evaluate its parameters, we insert it into the differential
equation. Using trigonometric identities such as $2\cos A \cos 2A=\cos A +\cos 3A$ to reduce
products to sums and dropping higher order terms, we obtain
\begin{eqnarray*}
 s\omega b+[s^s+\omega_0^2-(\omega/2)^2+(h/2)\omega_0^2]a=0,\\
s\omega a-[s^s+\omega_0^2-(\omega/2)^2-(h/2)\omega_0^2]b=0.
\end{eqnarray*}
For a near resonance solution we write $\omega =2\omega_0+\epsilon$ and neglect second order
terms to get
\begin{eqnarray*}
 sb-\frac{1}{2}[\epsilon-\frac{1}{2}h\omega_0]a=0,\\
sa+\frac{1}{2}[\epsilon+\frac{1}{2}h\omega_0]b=0.
\end{eqnarray*}
These equations can be solved for the coefficients provided
\begin{equation}
 s^2=\frac{1}{4}[(\frac{1}{2}h\omega_0)^2-\epsilon^2]>0.\label{61}
\end{equation}
Thus, we have a resonance with width
\begin{equation}
\Delta \omega=2\epsilon=h\omega_0=9.283\times10^{-3}\omega_0,\label{62}
\end{equation}
or
\begin{equation}
\Delta p=h(80.9\,\rm{MeV/c})=0.751\,\rm{MeV/c}.\label{63}
\end{equation}
And for the amplification factor at resonance we have
\begin{equation}
st=\frac{1}{4}h\omega_0t=\frac{\pi}{2}hn=1.46\times10^{-2}n,
\label{64}
\end{equation}
where $n$ is the number of atoms traversed in a resonant state. Since $\ln2=0.693$, this implies
that the amplitude is doubled in traversing about $50$ atoms.

Now note that the value of $h=\lambda_e/R_0$ used in (\ref{62}) and (\ref{64}) applies to only
the subclass of orbits for which $r_0=0.50 \AA$. For smaller radii the values can be much
larger. In principle, to get the width of the ensemble of orbits we should replace $h$ in
(\ref{62}) by its average $\obh$ over the ensemble. However, the result will probably not
differ much from the typical value we have chosen.

Similarly, the amplitude factor in (\ref{64}) will have a distribution of values, and the
doubling factor will be reached much faster for orbits with smaller radii.
Presumably, random perturbations (such as thermal fluctuations of the nuclei) will limit the
resonant state coherence length to some mean value $\overline{n}$. Consequently, states
with smaller $r_0$ will be preferentially ejected from the beam.

 Even more to the point, the perturbation parameter $h=\lambda_e/R$ is not constant as we
tentatively supposed but increases rapidly as the electron approaches a nucleus. In resonance
the value of $h$ close to each nucleus dominates the perturbation, so its effective mean value
is much smaller than the estimate for constant $r_0$.  Evidently, resonant interaction may
eject electrons with small $r_0$ in just a few atomic steps.

\subsubsection{Experimental implications}

The predicted resonance width in (\ref{63}) is in fair agreement with the width in the
channeling experiment data \cite{Gouanere05}, considering uncertainties in the value of $h$ and
such factors as thermal vibrations that may contribute to damping. Damping can only
narrow the width and destroy the resonance if it is too severe.

We need to explain how the orbital resonance is manifested in the experimental
measurements.   Two scintillators, SC2 and SC3, were employed to detect
the transmitted electrons. The larger detector SC3, with a radius about 3 times that of SC2,
served as a monitor while the smaller detector served as a counter for a central portion of
the beam. The measured quantity was the ratio of SC2 to SC3 counts as momentum was varied
in small steps $(0.083\%)$ over a 2\% range centered at the expected resonance momentum
$80.9$ MeV. An 8\% dip was observed at $81.1$ MeV.
Orbital resonances may contribute to this effect in at least two ways: first, and perhaps
most important, by increasing the probability of close encounter with a nucleus that will
scatter the electron out of the beam; second, by increasing the duration of eccentric orbits
outside the central region. Overall, resonant interactions will be strongest on electrons
confined to the central region.
These observations suffice for a qualitative explanation of the observed dip.
A quantitative calculation will not be attempted here.

The most problematic feature of the experiment is the $0.226\,\rm{MeV/c}$ difference
between observed and predicted resonance energies. If estimation of the experimental error was
overly pessimistic, that indicates a physical frequency shift. The most likely origin for such a
shift is the frequency split in (\ref{33}).  The experiment was not sufficiently accurate to
resolve separate peaks for the two frequencies, so the peaks would merge to broaden the
measured resonance width.  However, the peak for $\omega_+$ is likely to be higher than the
peak for $\omega_-$ owing to greater probability for ejection from the beam. Hence, the
center of the merged distribution will be displaced to a higher frequency. If this explanation is
correct, then an increase in experimental resolution will separate the two peaks, and their
relative heights will measure the relative probability of ejection at the two frequencies.

Though the string approximation to the crystal potential (\ref{23}) is useful for
semiquantitative analysis of zitter resonance, as we have seen, it breaks down completely at
radii near the screening radius $a=0.190\AA$; for then, as Lindhard \cite{Lindhard65} has shown,
the electron's collision time with a nucleus is comparable to the travel time between
atoms. In picturesque terms, the electron begins to ``feel'' individual atoms rather than a
continuous string. Within this domain, our analysis of electron motion remains qualitatively the
same, but a more realistic crystal potential is needed for accurate quantitative estimates.
A hard lower limit on the radius is determined by the mean radius of nuclear zeropoint
vibrations, which Debye theory estimates as $0.05 \AA$ \cite{Gemmell74}.

If the idea of zitter resonance is taken seriously, there are many opportunities for new
theoretical and experimental investigations.  Increasing the resolution by three orders of
magnitude will open the door to refined studies of frequency shifts, line splitting, spin
effects and Zeeman splitting, all of which are inherent in zitter theory \cite{Hest07}.
As has been noted, the most straightforward prediction of the zitter model is a second
order resonance near $161.7 \rm{MeV/c}$. In a first approximation, it can be analyzed in
much the same way as here, though removal of electrons from the center of the beam may be
enhanced by such processes as pair creation.

A more detailed analysis of zitter resonance in channeling requires close attention to
experimental conditions, so that will be addressed elsewhere.
Classical particle models have long been used for channeling calculations with considerable
success. Besides being simpler and more transparent than quantum mechanical models, they
even often give better results at high energies.
For present purposes, the zitter model differs from the usual classical model only by the
zitter dipole interaction.
As we see in the next Section, though zitter is inherent in the Dirac electron theory, it
is unlikely that the channeling resonance effect can be derived from the Dirac equation
without some modification such as projection into a Majorana state.

\subsection{Solution by separation of variables}

An obvious strategy for solving the rotor equation of motion is to separate  internal
zitter oscillations from the effect of external field on overall motion of the
comoving frame. Spinor equations admit a natural way to do that. For
a given electromagnetic field $F$, the rotor $R=R(\tau)$
can be factored into
\begin{equation}
 R=LU, \label{6.61}
\end{equation}
where rotor $L$ satisfies the equation
\begin{equation}
\dot{L}=\frac{q}{2m_e}FL.\label{6.62}
\end{equation}
Moreover, instead of using zitter averages as before, we can define a mean velocity $\obv$
for the zitter center by requiring

\begin{equation}
\obv=R\gamma_0\tR= L\gamma_0\tL \qquad \hbox{so} \qquad U\gamma_0\tU=\gamma_0.\label{6.63}
\end{equation}
This generalizes (\ref{6.14}) by dropping the requirement that L be a boost. However,
the boost condition placed no restrictions on equations for the velocity, whereas here we
have
\begin{equation}
m_e\dot{\obv}=qF\bdot \obv,\label{6.64}
\end{equation}
for which the relation to the momentum equation (\ref{4.33}), or even its zitter mean
(\ref{6.6}), is problematic in general. This drawback aside, the approach has sufficient
advantages to consider its use for approximate solutions at least.

Its first big advantage is evident in the transformed spin equation of motion
(\ref{6.24}), for which the $\Omega_0$ has been adjusted to cancel $F_0$, so it
reduces to
\begin{equation}
\dot{S}_0 =\frac{\hbar}{2}\frac{d}{d\tau}(\e_1+i\e_3)=m\e_2
.\label{6.65}
\end{equation}
This is equivalent to
the rotor equation
\begin{eqnarray}
&&\hspace{-.5cm}\dot{U}=\frac{1}{2}\omega i\,\e_3\,U=\frac{1}{2}Ui\bsig_3\omega\nonumber\\
&&\hspace{-.5cm}\hbox{with}\qquad \e_k=U\bsig_k\tU.
\label{6.66}
\end{eqnarray}
It has the obvious solution $U=U_0\exp{i\bsig_3\varphi}$, though $\omega=\dot{\varphi}$
must be obtained from other considerations.

A second big advantage is that exact solutions of (\ref{6.62}) have already been found for
several kinds of electromagnetic field \cite{Hest03b,Hest79}. Unfortunately, they all
presume that motion is along a timelike curve $x(\tau)$, whereas, in the present case
$F=F(z)=F(x+r)$ is given along the lightlike curve $z(\tau)$, so problems arise in
relating the curves $x(\tau)$ and $z(\tau)$.

Another difficulty is that the constraint equation relating zitter phase $\varphi$
to proper time is nonlinearly related to the rotor equation of motion, as is obvious
when expressed in the form
\begin{equation}
\dot{\varphi}=\omega_e+\frac{q}{m_e}<FLS_0\tL >.\label{6.67}
\end{equation}
Consequently, a closed solution of the coupled equations is not to be expected in most
cases. However, the interaction term in (\ref{6.67}) is invariably much smaller than the
constant term $\omega_e$, so it can often be neglected or, in any case, incorporated by
iteration with the equation of motion.

The complications just described disappear when $F$ is uniformly constant, as in the
case of any combination of homogeneous electric and magnetic fields.
Without working out the details, let's note some special features of this case that
can contribute to an elegant  solution.

For constant $F$, equation (\ref{6.62}) has the simple solution
\begin{equation}
L=\exp\left({\frac{q}{2m_e}F\tau}\right).\label{6.68}
\end{equation}
Since $FL=LF$ in this case, (\ref{6.67}) reduces to
\begin{equation}
\hspace{-.9cm}\dot{\varphi}=\omega_e+\frac{q}{m_e}<FS_0 >=\omega_e+\frac{q}{m_e}(\E\bdot \e_1
-\B\bdot\e_3).\label{6.69}
\end{equation}
The initial phase can be chosen to make the angle dependence of the zitter oscillations
explicit:
\begin{equation}
\dot{\varphi}=\omega_e-\frac{q}{m_e}\B\bdot\e_3+\frac{q}{m_e}|\E\btimes\e_3|\sin
\varphi.\label{6.70}
\end{equation}
This equation is akin to Kepler's equation in celestial mechanics relating incommensurate
time variables. Since the oscillating term is so small, solution by perturbation expansion
is quite satisfactory.

Furthermore, from the momentum equation of motion (\ref{4.33}) we find immediately that
$p-qF\bdot z$ is a constant of motion. And the equation of motion (\ref{4.39}) for
$\pi=p-m_e u$  reduces to
\begin{equation}
\dot{\pi}=\Omega\bdot\pi=-m\omega_e e_1,\label{6.70a}
\end{equation}
so $\pi^2=p^2-2m_e m$ is another constant of motion.

Finally, a nice trick for integration to get an explicit equation for the particle
history is given in \cite{Hest79}. This is all we need for the complete solution
of motion in arbitrary homogeneous fields. Note that complications of solving for the
wobbling momentum vector have been avoided.

\section{Zitter resonance and field theory}

Like the original Dirac theory, our zitter model describes a test particle, which is
to say that the electromagnetic field that it generates is ignored. But if we take the
zitter charge circulation seriously, it is evident that the particle must be the source
of a fluctuating electromagnetic field. The average of that field over a zitter period
will give us the static field of a magnetic dipole and an
effective Coulomb field with a virtual source at the zitter center.  Superimposed on these
mean values we have an electric dipole field rapidly rotating with the zitter
frequency.  This picture raises many questions that cannot be answered without
constructing a complete electromagnetic field theory with zitter sources. That is too much
to attempt here, but we can identify some of the issues and prospective physical
implications.

We consider here only a classical field theory, leaving issues of second quantization
and many particle theory for another day.
The first question is: What maintains the zitter charge circulation? Classical
electrodynamics implies that all accelerating charges radiate, which lead to collapse
of circulation.  An obvious suggestion is that circulation is maintained by self
interaction of the charge with its own field! Of course, that cannot be regarded as a
satisfactory answer until the self-interaction is calculated, and no such calculation has
ever been successful in any version of classical theory. Nevertheless, it does provide a
new constraint on the self-interaction problem, namely, that its solution should account
for electron spin as well as mass.   Besides that, there are tricky questions about
singularities in the field of a lightlike source.
Evidently some new idea will be needed to make zitter field theory work.
However, we can take an intermediate stance, presuming the zitter field as real and
exploring its implications.

The dipole interaction $\d\bdot \E$ is, of course, unobservable for laboratory electric
fields, because $\d$ rotates so rapidly. However, if $\E$ is the oscillating zitter field
of an electron, the possibility of resonance exists.
The following \textit{speculative survey} suggests that many surprising features of
quantum mechanics might be consequences of zitter resonances.

\textbf{Electron diffraction} \textit{as resonant momentum transfer mediated
by the zitter field.} An electron incident on a crystal is preceded by its zitter field which
is reflected back from a periodic array of scattering centers. At the Bragg incidence
angles, phase coherence of the reflected field oscillations with the
electron zitter provides a mechanism for resonant momentum transfer.
The propagating zitter field has a wavelength that may be identified with the electron
\textit{de Broglie wavelength} in quantum mechanics.

\textbf{Quantized atomic states} might be explained by the same
mechanism as diffraction, that is, as resonances of the electron zitter with its own field
reflected off the nucleus. This raises questions
about the topology of the zitter field that
may be related to \textit{Berry phase} and the \textit{Aharonov-Bohm effect.}
Each electron in a multi-electron atom will have its own zitter frequency, shifted by its
interaction energy in the atom. Consequently, an atom or molecule will have a zitter field
composed of the multiple frequencies of its constituents, and it will have not a
single de Broglie wavelength but a spectrum of wavelengths. Perhaps this spectrum will
show up in diffraction experiments.

The \textbf{Pauli Principle} \textit{as resonant phase locking.} In general, the
zitter oscillations of different electrons will not be coherent, because of
doppler shifts and frequency shifts due to local interactions. However, when two
electrons are in the same quantum state the necessary conditions for resonant phase
locking are fulfilled. It remains to be proved that antiparallel alignment of spins is
necessary for a stable state.

\textbf{Van der Waals and Casimir forces} are generally explained as
effects of coherent coupling between fluctuating dipoles. This involves coherent
coupling of phases, which is what zitter is all about. The zitter field serves as a mechanism
for maintaining that coupling that may have additional consequences. As the dipoles are
mutually induced, the interaction mechanism appears to be essentially the same as in
electron diffraction.

The \textbf{Lamb Shift} is commonly attributed to a smearing out of
electron position due to vacuum fluctuations. Obviously, the zitter offers an alternative
explanation for smearing out. Indeed, one can conceive an electromagnetic vacuum field
composed of a stochastic combination of zitter fields of all existing charged fermions.

Similar speculative explanations can be adduced for tunneling, anomalous magnetic
moment, covalent bonding and other important quantum effects.
Even Bose-Einstein condensates might be explained by resonant zitter coupling of
constituent fermions.
Zitter field theory must provide reasonable accounts of all these phenomena before it can
be  regarded as a viable extension of standard quantum mechanics.
However, we don't have to wait for that! If zitter is a real physical phenomenon we should be
able to devise experiments to observe it directly. In fact, as we have noted, that may
already have been done!

\section{Zitterbewegung in Dirac theory}

This section describes the intimate relation of the  zitter model to the Dirac
equation.  The relation should not be surprising, as it
was study of Dirac theory that led to the zitter model in the first place.
Conversely, we shall see that the zitter model suggests
modification of Dirac theory to incorporate deeper zitter substructure, with new physical
implications including a surprising possible connection to weak interactions.

Our first task is to match up variables and dynamical equations in the zitter model with
observables and dynamics in Dirac theory. An exact match is not to be expected, as the
particle based zitter model can only be related to the Dirac field theory by some sort of
averaging or projection process. Nevertheless, the comparison reveals specific similarities
and differences that must be addressed in establishing a firm connection between zitter model
and Dirac theory. This leads to suggestions for modifying the Dirac theory and further
research to resolve outstanding issues.

In the language of STA, the \textit{real Dirac equation} has the form
\begin{eqnarray}
\gamma^\mu\partial_\mu(\psi \gamma_2\gamma_1\hbar-qA_\mu\psi)&=&m_o\psi\gamma_0,
\nonumber\\
\hbox{or}\quad \nabla\psi i\bsig_3\hbar-qA\psi&=&m_e\psi\gamma_0\,,\label{7.1}
\end{eqnarray}
where the Dirac wave function is a   \textit{real spinor field}
\begin{equation}
\psi=\psi(x)=(\rho e^{i\beta})^{1/2}R. \label{7.2}
\end{equation}
This version of the Dirac equation is fully equivalent to the standard matrix
version \cite{Hest03b}, but it has great advantages for analyzing the structure of the
Dirac theory as shown in the following.

The Dirac wave function determines a frame field of \textit{local observables}
\begin{equation}
\psi\gamma_\mu\tilde{\psi}= \rho e_\mu ,
\quad \hbox{where} \quad
e_\mu =R\gamma_\mu \tilde{R}=e_\mu(x)  \label{7.3}
\end{equation}
and $\rho=\rho(x)$ is interpreted as a scalar probability density, in accordance with
the interpretation of the \textit{Dirac  current} $\psi\gamma_0\tilde{\psi}= \rho e_0$ as
a probability current. The vector fields
\begin{equation}
v\equiv e_0= R\gamma_0\tR=v(x),\label{7.4a}
\end{equation}
and
\begin{equation}
s\equiv \frac{1}{2}e_3 =\half \hbar R\gamma_3\tR=s(x) \label{7.4}
\end{equation}
are interpreted as local velocity and spin observables for the electron.
Note that these quantities are algebraically identical to the expressions
 for velocity and spin in the zitter model.
Likewise for the other observables
\begin{equation}
u\equiv e_0+e_2=R\gamma_+\tR=u(x) ,\label{7.4c}
\end{equation}
and
\begin{equation}
\obS \equiv isv=\half Ri\bsig_3 \hbar\tR =\obS(x).\label{7.4b}
\end{equation}
The difference is that the observables here are vector and bivector fields,
whereas in the zitter model they are defined on a particle history. Our next task is
to compare them dynamically.

\subsection{Zitterbewegung along Dirac histories}

The mass term in the Dirac equation (\ref{7.1}) can be written in the form
$$m_e\psi\gamma_0=m_e(\rho e^{i\beta})^{1/2}R\gamma_0\tR R=m_eve^{-i\beta}\psi.$$
Whence, the Dirac equation can be reformulated in the compact form
\begin{equation}
D\psi\equiv \gamma^\mu D_\mu\psi=0,\label{7.5}
\end{equation}
where a gauge invariant \textit{coderivative} is defined by
\begin{equation}
D_\mu=\partial_\mu+\half\Omega_\mu,\label{7.5a}
\end{equation}
with the bivector-valued connexion
\begin{equation}
\Omega_\mu\equiv \frac{2}{\hbar}(m_ev_\mu e^{-i\beta}+qA_\mu)e_2e_1.
\label{7.6}
\end{equation}
The purpose of introducing this coderivative is not to reduce the Dirac equation to the
maximally compact form (\ref{7.5}), but to reveal that all the essential physics is contained
in the connexion $\Omega_\mu$.

Our aim is to compare zitter dynamics along the history of a zitter center with Dirac
dynamics along ``streamlines" of the Dirac current.
To that end, we evaluate the
\textit{directional coderivative} along a Dirac history as follows. From
$$ D\psi\gamma_0= D(v
e^{-i\beta}\psi)=[D(v\psi)-i(\nabla\beta)v\psi]e^{-i\beta}=0$$
we obtain
\begin{equation}
D(v\psi)=2v\bdot D\psi +(Dv)\psi=i(\nabla\beta)v\psi.
  \label{7.7}
\end{equation}
This separates into the familiar conservation law for the Dirac current and a dynamical
equation for rotations along a Dirac streamline:
\begin{equation}
\hspace{-.8cm}\dot{\rho}+\rho D\bdot v=D\bdot(\rho v)=\nabla \bdot(\rho v)=0,\label{7.8}
\end{equation}
\begin{equation}
\hspace{-.8cm}v\bdot D R=\dot{R}+\half \Omega(v)R=-\half[D\wedge v+i(v\wedge\nabla\beta)]R.\label{7.10}
\end{equation}
The overdot indicates the directional derivative $v\bdot\nabla$ and (\ref{7.6}) gives us
\begin{equation}
\hspace{-.8cm}\Omega(v)=(\omega_0\cos\beta+ \frac{2q}{\hbar}A\bdot v)e_2e_1
+(\omega_0\sin\beta )e_0e_3\label{7.10a}
\end{equation}
\begin{equation}
\hspace{-.8cm}D\wedge v=\nabla\wedge v+(\omega_0\sin\beta )e_3 e_0.\label{7.11}
\end{equation}
When these are inserted into (\ref{7.10}), the boosts in the $e_0e_3$ plane cancel to give us
\begin{equation}
\hspace{-.9cm}\dot{R}=\half[(\omega_0\cos\beta+ \frac{2q}{\hbar}A\bdot v)e_1e_2-\nabla\wedge
v-i(v\wedge\nabla\beta)]R.\label{7.11a}
\end{equation}
This is an exact result. It does indeed exhibit the familiar zitter rotation in the
$e_2e_1$-plane, though the frequency seems to be different.

We still need to evaluate the curl of the velocity field to appreciate its effect on the
dynamics. A general
expression has been derived elsewhere \cite{Hest03c,Hest96}, but our purpose here is
served by the eikonal approximation, expressed by
\begin{equation}
\hspace{-.4cm}\nabla \psi \gamma_2\gamma_1\hbar=\nabla (\psi_0e^{-\gamma_2\gamma_1\Phi/\hbar}
\gamma_2\gamma_1\hbar) =(\nabla\Phi)\psi.\label{7.12}
\end{equation}
Inserting this into the Dirac equation (\ref{7.1}), we obtain
\begin{equation}
\nabla\Phi=m_eve^{-i\beta}-qA.  \label{7.13}
\end{equation}
This implies $e^{i\beta}=\pm 1$,
where the choice of sign depends on the chosen sign of charge.
We adopt that approximation only in this equation, as the parameter $\beta$ is too
important to ignore completely. Then the curl of (\ref{7.13}) gives us
\begin{equation}
-m_e\nabla\wedge v=q\nabla\wedge A=qF.\label{7.14}
\end{equation}
This is equivalent to the Lorentz force equation for a fluid of charge with uniform
density, as seen by ``dotting" with $v$, to get
\begin{equation}
\hspace{-.4cm}v\bdot(\nabla\wedge v)=v\bdot\nabla v=-\frac{q}{m_e}v\bdot F
=\frac{q}{m_e} F\bdot v.\label{7.15}
\end{equation}
More generally, we insert (\ref{7.14}) into (\ref{7.10}) and use (\ref{7.11}) to get
the spinor equation of motion
\begin{equation}
\dot{R}= \half [\omega_v e_1e_2+\frac{q}{m_e} F
-i(v\wedge\nabla\beta)]R, \label{7.16}
\end{equation}
with $\omega_v\equiv\omega_0\cos\beta+ (2q/\hbar)A\bdot v$. This equation must be
compared with the analogous expression (\ref{6.9}) for rotational velocity in the  minimal
model. The general form is very similar, but we are not equipped to account for the
differences for reasons to be discussed.

The zitter frequencies in the two equations are not equivalent, but they do
have the same free particle limit $\omega_0$. The bivector
$i(v\wedge\nabla\beta)$ also contributes to the rotation rate, as implied by
$$v\bdot[i(v\wedge\nabla\beta)]=-i(v\wedge v\wedge\nabla\beta)=0.$$
However, the physical significance of this term remains obscure.
The apparent absence of a Stern-Gerlach force in the
Dirac version (\ref{7.16}) is noteworthy, but we cannot be sure that it is not buried in
terms that we do not understand.

It might be thought that the Dirac equation is more fundamental than the zitter model
because interaction comes from the vector potential $A$ alone and interaction with the field
$F$ arises only indirectly, for example in the manner described above. A famous consequence
of this is the derivation of the gyromagnetic ratio $g=2$. In contrast, the Lagrangian
(\ref{4.1}) for the zitter model appears to presume the electron magnetic moment with
separate coupling constants for A and F interactions. Note, however, that precisely
two independent constants are presumed in both models. The rest mass is presumed in the
Dirac equation, but that is replaced by the zitter radius in the zitter model. It remains
to be seen which is more fundamental.

\subsection{Zitterbewegung substructure}
\smallskip

In preceding sections we saw that a rotating electric dipole is the hallmark of
zitterbewegung, so one wonders why it has attracted so little attention in accounts
of Dirac theory.  In the original paper introducing his equation
\cite{Dirac28}, Dirac concluded that the electron has both a magnetic and an electric
moment, the magnetic moment being the same as in the Pauli model. However, he said,
``The electric moment, being a pure imaginary, we should not expect to appear in the
model. It is doubtful whether the electric moment has any physical meaning."

It is worth translating Dirac's argument into STA, especially since its mathematical
content has been retained in the current literature.
One simply ``squares" the operator on the wave function in (\ref{7.1}) to get
\begin{equation}
\hspace{-.9cm}(-\hbar^2\nabla^2+q^2A^2)\psi
-2q(A\bdot\nabla+\half F)\psi i\bsig_3\hbar=m_e^2\psi\,,\label{7.17}
\end{equation}
where $F=\nabla A=\nabla\wedge A$ with the \textit{Lorenz condition} $\nabla\bdot A=0$.
This is the Klein-Gordon equation with an extra term that explicitly shows the action
of bivector $F$ ``rotating" the wave function. The interaction energy density associated with
this term is proportional to
\begin{multline}
\half< F\psi i\bsig_3\hbar\tpsi>=< F\obS \rho e^{i\beta}>\\
=-\rho (\B_v \bdot \s\, \cos \beta+\E_v\bdot\s\, \sin \beta),\label{7.18}
\end{multline}
where the $v$-split introduced in (\ref{6.12}) has been used on the right hand side.
The $\B_v \bdot \s\, \cos \beta$ term is recognized as the Pauli term except for the
strange $\cos\beta$ factor. The $\E_v\bdot\s\, \sin \beta$ term is what Dirac identified
as an imaginary dipole moment.
He never mentioned the electric dipole again. In his influential
textbook \cite{Dirac58} he simply suppressed the offending term by a subterfuge
advertised as a change in representation. Then he killed the
term with an approximation that amounts to $ \sin \beta=0$ and never looked
back.
We find an alternative resolution of this ``dipole problem" below.

Dirac was soon convinced by Schroedinger \cite{Schr30} that zitterbewegung is foundational
to electron theory and he argued the case vigorously in his textbook \cite{Dirac58}.
As his argument is still widely accepted \cite{Greiner90}, it deserves comment here.

Dirac introduces a position operator by identifying a velocity operator as its time
derivative, and he followed Schroedinger in integrating the equation for the free
particle case.
He identifies his $\alpha_k$ matrices as velocity operators and claims that their
eigenvalues $\pm 1$ correspond to measured values of electron velocity, asserting:
``we can conclude  that \textit{a measurement of a component of the velocity of a free
electron is certain to lead to the result $\pm c$.}" (Dirac's italics)
From the STA point of view this argument and its implications are bogus, for reasons
explained elsewhere \cite{Hest03b}.
However, we agree with the assumption that the electron moves with the
speed of light.  We differ in identifying the local electron velocity with the
null vector $u=R\gamma_+\tR$ introduced above.

We also agree with Dirac in attributing the origin of spin to zitter. Dirac concludes:
`` Our argument is valid only provided the position of the particle is an observable.
If this assumption holds, the particle must have a spin angular momentum of half a
quantum."
However, Dirac's analysis of zitter and spin never went beyond the free particle case.
He overlooked (or dismissed) the inference that his account of spin arising from
charged particle circulation implies a rapidly rotating electric dipole.
Though zitter obviously arose from wave function phase oscillations in Schroedinger's free
particle analysis, Dirac never considered a general connection of zitter circulation to
wave function phase.
The present study can be regarded as an extension of Dirac's analysis to incorporate
these features in a general theory of zitter in quantum mechanics.

As the Dirac equation has an unsurpassed record of success in QM and QED
applications, it is imperative to reconcile it with any proposals about zitter.
If the zitter model describes  substructure in electron motion that is not
captured by the Dirac equation, it must at least be related to the Dirac wave function
by some sort of averaging process.
Without attempting a definitive reconciliation, let us note some issues that must be
addressed.

We know that the conserved Dirac current $\rho v$ determines a congruence of curves (or
streamlines) for every solution of the Dirac equation.
As Bohm and Hiley have argued at length \cite{Bohm93}, each of these curves $x=x(\tau)$
can be regarded as a possible path for the electron weighted by a relative probability
$\rho=\rho(x)$ that the electron actually followed that path. This is a viable
particle interpretation of quantum mechanics.
However, a refinement is necessary to account for zitterbewegung, which suggests that
the actual particle paths are lightlike helices with tangent vector $u=u(x)$ at each
spacetime point.
The simplest refinement would have each of these lightlike paths winding around a
Dirac streamline, but this possibility is questionable without deriving it from the
zitter model, at least approximately, by a well defined averaging
process (yet to be determined)!

A crucial problem is to justify the weighting of paths by the probability density
$\rho=\rho(x)$.
A new possibility is suggested by the fact that the analogous quantity in our
zitter particle model  is a timescale factor
$\rho=\rho(\varphi)=\frac{d\tau}{d\varphi}$, which, as we have noted, would arise naturally
from averaging over paths with different proper times but a common phase angle $\varphi$.
Thus, the putative probability density in the Dirac equation might be derivable as a time
scale weighting on a congruence of particle paths! In any case, time scaling in the zitter
model must be reconciled with the probability interpretation in the Dirac theory.

It seems likely that a suitable averaging process relating particle histories to the Dirac
equation will involve time averaging as well as ensemble averaging with constraints.
Analogy with the zitter model already suggests that  Dirac observables for velocity $v$ and
spin $\obS$ correspond to zitter time averages, but the phase of the Dirac wave function
is directly comparable to the zitter phase.

The Schroedinger equation is a nonrelativistic approximation to the Dirac equation that
freezes spin but preserves zitter oscillations in the phase of the wave function and in
coupling of phase to amplitude, as specified, surprisingly, by the mysterious
parameter $\beta$ \cite{Hest79,Recami98}.
As shown in these references, assuming $\beta=0$ completely decouples phase from
amplitude and so eliminates all QM effects.
Clearly, therefore, the role of $\beta$ must be accounted for in any averaging
process.

The fact that the kinematic state of a particle with zitter is described by a
rotor  while the Dirac wavefunction is also a spinor suggests that the
QM \textit{superposition principle} can be construed as an average over particle
rotor states.
The undeniable success of the Schroedinger equation suggests that, approximately at
least, the superposition is an average over phase factors, such as Feynman's sum over
paths. This certainly produces the coupling of phase to amplitude so characteristic of
QM.
Note that  an average over rotors blurs any average over paths, because path
velocity is a bilinear function of the path rotor.
Finally, it should be mentioned that gauge invariance provides a strong constraint on
assembly of phases for distinct particle paths into a coherent ensemble.

Clearly, deriving the Dirac equation from zitter substructure is a
nontrivial problem.  Happily, we don't have to wait for a solution to
make progress in studying the zitterbewegung. Here is a promising alternative approach:

\subsection{Putting zitterbewegung into the Dirac equation}

We have seen that physical interpretation of the Dirac equation is crucially dependent
on identification of a particle velocity observable, which requires theoretical
assumptions beyond the Dirac equation itself.
Historically, the mass term in the Dirac equation led to the conservation
law for the Dirac current and its interpretation as a probability current, with an
implicit identification of particle velocity.
In the STA version of the Dirac equation (\ref{7.1}),  explicit appearance of the
vector $\gamma_0$ shows that the velocity vector $v=R\gamma_0\tR$ is inherent in the
structure of the equation.
However, we have identified the vector $u=R\gamma_+\tR$ as a better candidate for
electron velocity. This suggests a slight modification of the Dirac equation to
replace $v$ by $u$. Accordingly, we change the mass term $m_e\psi\gamma_0$ to
\begin{equation}
m_e\psi\half (\gamma_0+\gamma_2)=m_e\psi_+\gamma_0,\label{7.19}
\end{equation}
where
\begin{equation}
\psi_+\equiv \psi\half (1+\bsig_2).\label{7.20}
\end{equation}
In common parlance, this is a projection of a 4-component spinor $\psi$ into a
2-component spinor $\psi_+$. As only these components are now relevant to the electron
velocity, we should perform the same projection on the other components of the Dirac
equation. That requires modification of the vector potential term to achieve a
2-component equation.
Thus we arrive at the modified Dirac equation
\begin{equation}
\nabla\psi_+ i\bsig_3\hbar-qA\psi_+\bsig_3=m_e\psi_+\gamma_0\,.\label{7.21}
\end{equation}
Let's refer to this as the \textit{zitter Dirac equation}.
Note that it is invariant under projection from the right by
$\half (1-\bsig_2)$.

To be assured that this modification of the Dirac equation has not damaged its essential
physical meaning, we note that the zitter free particle solution (\ref{5.13a}) is also a
solution of the Dirac equation, which, when substituted into (\ref{7.21})  with
arbitrary initial conditions and $\i=i\bsig_3$, yields the algebraic relation
\begin{equation}
pu=m_e(1-e_2e_0).\label{7.21a}
\end{equation}
This is identical to the relation between momentum and velocity found in (\ref{4.19}) for
the zitter model. In contrast to the prosaic relation $p=m_ee_0$ from the free
particle solution to the ordinary Dirac equation, this relation includes zitter in the
 vector $e_2$, as it rotates with the zitter phase. Thus, the form of the zitter Dirac
equation brings the null velocity observable to the fore.

It is also readily shown that the zitter Dirac equation has the same electromagnetic gauge
invariance as the ordinary Dirac equation, though we shall see that the gauge generator assumes a
different form when the gauge group is generalized to incorporate electroweak interactions.

Now check the observables.
With respect to the projected wave function, the observables (\ref{7.4c}), (\ref{7.4b}) become
\begin{equation}
\psi_+\gamma_+\tpsi_+=2\psi_+\gamma_0\tpsi_+=\psi\gamma_+\tpsi=\rho u ,\label{7.22}
\end{equation}
and
\begin{multline}
\psi_+i\bsig_3 \hbar\tpsi_+=\psi\half (1+\bsig_2)i\bsig_3 \hbar\psi\\
=\frac{\hbar}{2}\psi \gamma_+\gamma_1\psi=\rho \frac{\hbar}{2}ue^{-i\beta}e_1
=\rho S.\label{7.23}
\end{multline}
Thus we get the same zitter velocity $u$, but the spin bivector ${\obS}$ is replaced by a
null spin bivector $S$, which is identical in form to the spin
(\ref{4.14}) in the point particle model, and, as in that case, the duality factor can
be absorbed into a rotation of the vector $e_1$.
We see immediately that the interaction energy density (\ref{7.18}) becomes
\begin{multline}
\half< F\psi_+ i\bsig_3\hbar\tpsi_+>=< F \rho S>\\
=\rho (\E_v\bdot\d-\B_v \bdot \s),\label{7.24}
\end{multline}
in perfect accord with equation (\ref{6.13}) of the zitter model. Thus, the change in
observables by projection on the wave function appears to eliminate the parameter $\beta$
and its problems of physical interpretation.

Further insight comes from the following Lagrangian for the zitter Dirac equation:
\begin{multline}
 L_{zD}=2\left<[-\hbar\nabla\psi_+i\gamma_0\phantom{q\tpsi}\right.\\ \left.
+qA\psi_+\gamma_0+m_e \psi_+\bsig_3]\tpsi\right>. \label{7.24a}
\end{multline}
Note that the interaction term has the usual form $ A\bdot J=<AJ>$, but now the charge
current is a null vector field
\begin{equation}
J=2q\psi_+\gamma_0\tpsi=q\psi\gamma_+\tpsi=q\rho u, \label{7.24b}
\end{equation}
as expected. It follows from the zitter Dirac equation that this current is conserved, though
it is not conserved in standard Dirac theory.

The usual Dirac current is not obtainable as a bilinear observable of the wave
function $\psi_+$. However, as noted before, it can be obtained as a zitter
average $v=\obu$ of the zitter velocity. This suggests that one should try to derive the
zitter Dirac equation, rather than the Dirac equation itself, from the zitter model.
But that possibility will not be explored here.

The zitter Dirac equation offers a new perspective on the significance of negative energy
in Dirac theory.
Recall that negative energy solutions were first regarded as a serious defect of the
Dirac equation.
Schroedinger showed that they are essential for a general solution even in the free
particle case when he ``discovered zitterbewegung" as interference between positive
and negative components of a wave packet.
In an audacious effort to save the theory, Dirac identified the negative energy states
with an ``anti-electron" and invoked the Pauli principle to suppress them (hole
theory).
Miraculously, the positron was discovered shortly thereafter, so the defect was
transmuted to a spectacular triumph!

A standard conclusion from all this is expressed by the following
quotations \cite{Greiner90}:
``The zitterbewegung  demonstrates in a real sense a single particle theory
is not possible."
``The difficulties with the negative energy states of the Dirac equation almost of
necessity demand a many-body theory."
``\textit{Hole theory is a many-body theory} describing particles with positive and
negative charge. The simple probability interpretations of the wave functions
acclaimed in a single-particle theory cannot be true any longer, because the creation
and annihilation of electron-positron pairs must be taken in account in the wave
function."
In other words, quantum field theory is needed to explain zitterbewegung!

For a different perspective on negative energy we note that $\psi^C\equiv\psi\bsig_2$
is the charge conjugate solution of the real Dirac equation \cite{Hest03b}.
Hence we can cast the zitter wave function (\ref{7.20}) in the form
\begin{equation}
\psi_+\equiv \half (\psi+\psi^C).\label{7.25}
\end{equation}
This expresses Dirac's negative energy solution as an essential component of the zitter
rather than an antiparticle. It is an alternative splitting of Dirac's 4-component
wave function into a pair of 2-component wave functions for different particle
states.
The physical issue is this: Which components of the Dirac wave function should be
identified with the electron?  The zitter component $\psi_+$ describes an electron with
zitter motion.  Standard quantum field theory splits the zitter into positive and negative
energy components and then reassembles it later from pair creation and annihilation.
In most calculations the end result will be the same, because both approaches start from
the same Dirac equation.
Contrary to the standard Dirac equation (\ref{7.1}), the zitter Dirac equation
(\ref{7.21}) is consistent with a single particle model of zitter without the strenuous
expedient of field quantization.
This is not to deny that some version of quantum field theory is necessary to account
for creation and annihilation of particles.
The problem is to devise experiments that identify the basic particle states.

\subsection{Electroweak interactions}

We have seen that only half the Dirac wave function $\psi_+= \psi\half
(1+\bsig_2)$ is needed to describe the electron. What can be said about the other half
$\psi_-= \psi\half(1-\bsig_2)$?
An attractive answer is suggested by gauge theory.

In the real Dirac equation (\ref{7.1}) an electromagnetic gauge transformation of the
wave function is multiplicative on the right with the form
\begin{equation}
\psi\quad\rightarrow\quad  \psi'= \psi U,\label{7.26}
\end{equation}
where $U=\exp{(i\bsig_3 \chi)}$.
We look to adapt this transformation to the zitter Dirac equation
(\ref{7.21}) and generalize it in a way that preserves essential structure of the
equation. To preserve the mass term, we require
\begin{equation}
\tilde{U}\gamma_0 U = \gamma_0. \label{7.27}
\end{equation}
The general solution of this equation has the form
\begin{equation}
U=e^{\frac{1}{2}i\,\btheta} e^{\frac{1}{2}i\,\chi}, \label{7.28}
\end{equation}
where $\btheta=\theta_1 \bsig_1+\theta_2 \bsig_2+\theta_3 \bsig_3$. Remarkably, this is
the gauge group SU(2)$\otimes$U(1) of electroweak theory. It strongly suggests that
the  geometric structure of electroweak theory is already inherent in Dirac theory!
Indeed, it requires that we identify $\psi_-\equiv\psi_\nu$ with the neutrino, just as
we identify $\psi_+\equiv\psi_e$ with the electron.

Thus, we construe the Dirac wave function as a lepton wave function composed of
electron and neutrino components:
\begin{equation}
\psi = \psi\half(1+\bsig_2) +\psi\half(1-\bsig_2)=\psi_e+\psi_\nu.\label{7.29}
\end{equation}
To adapt the Dirac equation to this interpretation, we follow standard electroweak
theory in introducing a gauge covariant derivative
\begin{eqnarray}
\hspace{-.4cm}&&\mathcal{D}_\mu\psi\equiv\partial_\mu\psi-\psi iW_\mu\nonumber\\
\hspace{-.4cm}&&\hbox{with} \qquad
W_\mu=\frac{1}{2\hbar}( g\mathbf{A}_\mu-g'B_\mu).\label{7.30}
\end{eqnarray}
In the standard way, gauge covariance is assured by requiring that the electroweak
connexion satisfies the transformation law
\begin{equation}
W_\mu\quad\rightarrow\quad  W_\mu'= \tilde{U}W_\mu U-\tilde{U} \partial_\mu U.
\label{7.31}
\end{equation}
Accordingly, the Dirac equation generalizes to the lepton wave equation
\begin{equation}
\gamma^\mu \mathcal{D}_\mu \psi i\bsig_3 \hbar=m_e\psi_e \gamma_0.\label{7.32}
\end{equation}
For the electron component, this reduces to the zitter Dirac equation (\ref{7.21}) when
weak interactions are turned off.

Details of this model integrating zitterbewegung with electroweak theory are discussed
elsewhere \cite{Hest07}.
It is amusing to recall that one of the early suggestions to account for extra
components in the Dirac wave function was to identify them as proton states.
Long dismissed because the particles had different mass, the idea returns again as an
electron-neutrino wave function.

\section{Conclusions}

The zitterbewegung, if it turns out to be physically real, is belated confirmation of de
Broglie's original hypothesis \cite{Broglie23} that the electron has an internal clock with
period precisely equal to twice the zitter period, precisely the relation between the period of
a rotor and that of a vector it rotates.

As we have seen, the physical signature of zitter is a rotating electric dipole with ultra
high frequency. If this exists, its implications for quantum mechanics will be far-reaching.
Evidently it can be incorporated in Dirac theory by subtle changes in the specification of
observables and the structure of the Dirac equation.

Experimental confirmation of the zitter should stimulate research on its proposed
incorporation into electroweak theory. Then study of zitter self-interaction should look for
excitations explaining the three lepton families. Finally, the strong analogy between
electroweak interactions of leptons and quarks suggests that one should investigate
modifications of zitter structure to model quarks and strong interactions. All this should
go hand-in-hand with development of zitter field theory and reconciliation of it with
quantum field theory.

\begin{acknowledgments}I am indebted to Richard Clawson for insightful observations about
analysis of the zitter model and to Michel Gouan\`ere for
explaining details of his channeling experiment and data analysis.
\end{acknowledgments}

\vskip.1in
\textbf{Note.} Many of the papers listed in the references are available online at
$<$http://modelingnts.la.asu.edu$>$ or $<$http://www.mrao.cam.ac.uk/$\sim$clifford/$>$.

\bibliography{zbw}

\end{document}